\documentclass[aps,prb,twocolumn,amsmath,amssymb,superscriptaddress,scrartcl,eqsecnum,bibliography]{revtex4-1}
\usepackage{amsmath,amsfonts,amssymb,graphics,graphicx,epsfig,color,times,indentfirst,layout}
\usepackage{extarrows}
\usepackage{slashed}
\usepackage{lipsum}
\usepackage{extarrows}
\usepackage{slashed}
\usepackage{lipsum}

\usepackage{mathrsfs}
\usepackage[unicode=true,pdfusetitle,bookmarks=false,colorlinks=true,citecolor=black,urlcolor=black,linkcolor=black]{hyperref}
\usepackage{tikz}
\usepackage{tikz-cd}
\usetikzlibrary{arrows}
\usetikzlibrary{intersections}
\usetikzlibrary{shapes.geometric}

\usepackage[T1]{fontenc}
\usepackage{bbold}

\def\bea{\begin{eqnarray}}
\def\eea{\end{eqnarray}}

\def\be{\begin{equation}}
\def\ee{\end{equation}}

\newcommand{\ket}[1]{|#1\rangle}

\newcommand{\lag}{\mathcal{L}}

\newcommand{\hilb}{\mathcal{H}}

\begin{document}

\title{Entanglement entropy of (3+1)D topological orders with excitations}

\date{\today}

\author{Xueda Wen}
\affiliation{Department of Physics, Massachusetts Institute of Technology, Cambridge, MA 02139}

\author{Huan He}
\affiliation{Physics Department, Princeton University, Princeton, New Jersey 08544, USA}

\author{Apoorv Tiwari}
\affiliation{Department of Physics and Institute for Condensed Matter Theory,
University of Illinois, 1110 W. Green Street, Urbana, IL 61801, USA}
\affiliation{Perimeter Institute for Theoretical Physics, Waterloo, Ontario N2L 2Y5, Canada}

\author{Yunqin Zheng}
\affiliation{Physics Department, Princeton University, Princeton, New Jersey 08544, USA}

\author{Peng Ye}
\affiliation{Department of Physics and Institute for Condensed Matter Theory,
University of Illinois, 1110 W. Green Street, Urbana, IL 61801, USA}

\begin{abstract}
Excitations in (3+1)D topologically ordered phases have very rich structures. (3+1)D topological phases support both point-like and string-like excitations, and in particular the loop (closed string) excitations may admit knotted and linked structures. In this work, we ask the question how different types of topological excitations contribute to the entanglement entropy, or alternatively, can we use the entanglement entropy to detect the structure of excitations, and further obtain the information of the underlying topological orders? We are mainly interested in (3+1)D topological orders that can be realized in Dijkgraaf-Witten gauge theories, which are labeled by a finite group $G$ and its group 4-cocycle $\omega\in\mathcal{H}^4[G;U(1)]$ up to group automorphisms. We find that each topological excitation contributes a universal constant $\ln d_i$ to the entanglement entropy, where $d_i$ is the quantum dimension that depends on both the structure of the excitation and the data $(G,\,\omega)$. The entanglement entropy of the excitations of the linked/unlinked topology can capture different information of the DW theory $(G,\,\omega)$. In particular, the entanglement entropy introduced by Hopf-link loop excitations can distinguish certain group 4-cocycles $\omega$ from the others.
\end{abstract}
\maketitle

\section{Introduction}
\label{Introduction}

The long range entanglement of topologically ordered phases in (2+1)D are closely related with the exotic features such as
fractional quasi-particle statistics and topologically protected ground state degeneracies (GSD)\cite{Wen04book,Wilczek1984,Wen1990,levin2006,kitaev2006,Li2008Entanglement,Dong0802,zhang2012, Regnault2009Topological,Thomale2010Nonlocal,Thomale2010Entanglement,Prodan2010Entanglement,Sterdyniak2011Extracting,Papic2011Topological,Hermanns2011Haldane,Chandran2011Bulk,Regnault2011Fractional,Sterdyniak2011Hierarchical,Alexandradinata2011Trace,Sterdyniak2012Real,Gilbert2012Signature,Sterdyniak2013Series,Estienne2015Correlation,Wen2016Edge}. By studying the topological entanglement entropy\cite{levin2006,kitaev2006} (TEE) of topological phases on non-trivial spatial manifolds such as a $T^2$, we can extract the universal topological data, for instance the modular $\mathcal{S}$, $\mathcal{T}$ matrices, of the emergent topological quantum field theory (TQFT)\cite{Dong0802,zhang2012,He2014Modular,Moradi1404,Mei2015Modular,Zhang2015General,you2015measuring,
Huang2015Transition,Huang2016Detecting,Li2017Experimental,Mei2017Gapped}.

Topological phases in (3+1)D have attracted extensive attention recently\cite{Walker2012,Curt2013,
Wang2014Braiding,Jiang2014Generalized,Moradi1404,JW1404,Wan1409,Bi2014Anyon,Wang2015Topological,Lin2015Loop,
Jian2014, Yoshida2015Topological,Chen2016Bulk,Wang2016Quantum,Wang2016Aspects,Tiwari2017Wilson,Wang2017Twisted,
Williamson2017311,
Chan2017Borromean,Cheng1511,Cheng2017Loop,Putrov2017,Chen2017,Else2017,DelCamp1709}. One of the demonstrative examples of topological phases in (3+1)D is the gauged symmetry protected topological phases (gSPT)\cite{Levin2012Braiding,Wang2014Braiding,Ye2013,Ye2016,He1608,Chan2017Borromean}. The symmetry protected topological (SPT) phases are topologically nontrivial gapped phases of matter protected by global symmetries\cite{Gu2009Tensor,Pollmann2010Entanglement,Chen2011Classification,Liu2011Gapped,Chen2011Complete,
Liu2011Symmetry,Pollmann2012Symmetry,Wen2012Symmetry,Tang2012Interacting,Liu2012Symmetry,Chen2012Chiral,
chen2012symmetry,ChenGuLiuWen,Liu2013Symmetry,Wen2013Classification,
ChenBurnell2015,AshvinSenthil,WangSenthil2013,Wang629,Metlitski2015,Metlitski2014arXiv,
Gu2014Symmetry,Ye2014Constructing,
Liu2014Microscopic,Wen2015Construction,Lan2017Classification, Tiwari:2017wqf}. The global symmetries of SPT phases can be gauged to obtain gSPT phases which support fractional excitations. The underlying TQFT describing the gSPT is the Dijkgraaf-Witten (DW) gauge theory\cite{DW1990}, which is characterized by a pair $(G,\,\omega)$ up to group automorphisms, where $G$ is a finite group and $\omega\in\mathcal{H}^4(G;U(1))$ is its group 4-cocycle. For the trivial cocycle $\omega$, namely an identity function from $(G)^4$ to $U(1)$, the DW theory is
called ``untwisted", while for nontrivial cocycles, the DW theory is called ``twisted".

\begin{figure}[t]
	\includegraphics[width=2.80in]{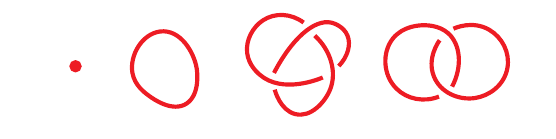}
	\caption{
		Examples of point-like and string-like excitations in (3+1)D topologically ordered phases.
		From left to right: A point excitation; a loop (or closed string) excitation;
		a trefoil-knot loop excitation; and a Hopf-link loop excitation.
	}
	\label{Excitations}
\end{figure}

Most recently, it has been proposed that the (3+1)D topological orders whose point-like excitations are all bosons
can be classified by unitary pointed fusion 2-categories\cite{LanKongWen2017}.
In particular, all of such (3+1)D topological orders
can be realized by DW gauge theories. Throughout this work, we are interested in
the (twisted/untwisted) DW theories.
The excitations of DW theories in three spatial dimensions include point-like excitations and string-like excitations
with possible knot and link topologies as depicted in Fig.~\ref{Excitations}.

The ground-state entanglement entropy of an untwisted DW theory with a general discrete gauge group $G$ on $R^3$ has been studied based on Kitaev's quantum double in Refs.~\onlinecite{Grover1108,Zheng2017Structure}. The entanglement entropy with respect to a $T^2$ entanglement cut is:
$
S=V\ln|G|-\gamma,
$
where
\begin{equation}\label{Gamma}
\gamma=\ln |G|
\end{equation}
is the TEE, $|G|$ is the order of gauge group $G$, and $V$ is the number of vertices on the boundary of the subregion which counts the area part of the entanglement entropy.

In this paper, aside from the ground-state entanglement entropy,
we study the effects of various excitations on the entanglement entropy in (3+1)D DW theories, in order to detect $(G,\,\omega)$ that characterizes a DW theory.
To calculate the contributions of the excitations to the entanglement entropies, we utilize the correspondence between minimal entangled states (MESs) and excitations\cite{zhang2012}.
It is known that MESs can be viewed as the bases for the degenerate ground states on a $d$-torus and correspond to different excitations by cutting the $d$-torus (see Fig.~\ref{MES} for the case of $T^3$). The MESs
may also be used to construct the modular $\mathcal{S}$ and $\mathcal{T}$ matrices. See, \textit{e.g.}, Refs.~\onlinecite{zhang2012,Hu2012} for (2+1)D and Refs.~\onlinecite{Jiang2014Generalized,DelCamp1709,Wan1409} for (3+1)D topologically ordered phases.

Based on this MES/excitation correspondence, we find that each topological excitation contributes to the entanglement
entropy with a universal constant
\be
\Delta S=\ln d_i,
\ee
where $d_i$ is the quantum dimension of the topological excitation. It depends on both the topologies of the excitations and $(G,\,\omega)$. In this paper, we mainly focus on three types of excitations in a DW theory, \text{\textit{i.e.}}, point-like excitations, single-loop excitations, and Hopf-link loop excitations, as interpreted in the following.

(i) Point-like excitations

Point-like excitations can be created either by local operators, or by non-local open string operators (or Wilson line operators). The former are called local point-like excitations and the latter are called topological point-like excitations, and in this work we are mainly interested in the latter case. Two topological point-like excitations are considered to be topologically equivalent if they differ by adding/removing local point-like excitations. As discussed in Ref.~\onlinecite{LanKongWen2017}, the point-like excitations in (3+1)D topologically ordered phases are characterized by a symmetric fusion category (SFC), which describes a collection of particles with trivial mutual
statistics. Depending on whether there are quasiparticles with fermionic statistics or not,
the point-like excitations are described by $\mathrm{sRep}(G)$ or $\mathrm{Rep}(G)$, respectively.
\footnote{Mathematically, $\mathrm{sRep}(G)$ is a category formed by the representations of $G$,
with some of the irreducible representations assigned bosonic statistics, while the others assigned fermi statistics.
$\mathrm{Rep}(G)$ is a category formed by the representations of $G$, with all the irreducible representations
assigned bosonic statistics. See, \textit{e.g.}, Ref.\onlinecite{LanKongWen2016} for more details.
}
For the DW theory, there are only bosonic point-like excitations which are described by $\mathrm{Rep}(G)$. Then the quantum dimension of a point-like excitation is
\begin{equation}\label{d_i}
d_i=\mathrm{dim}\left[\mathrm{Rep}_i(G)\right].
\end{equation}

(ii) Single-loop excitations

A single-loop excitation here refers to a loop excitation without any knotted structures.
It can be created at the boundary of a membrane operator on a disk.
 Each single-loop excitation is characterized by one flux (conjugacy class $\chi$)  and one charge (irreducible representation
of the centralizer $G_{\chi}$ with respect to $\chi$)\cite{Preskill199050,Preskill1992,Preskill1992NPB,LanKongWen2017}.
That is, one may use a pair $[\chi,\mathrm{Rep}(G_{\chi})]$ to label each single-loop excitation.
For a pure loop excitation which carries no charge, it is in the trivial irreducible representation of $G_{\chi}$.
From the group theory point of view, it is known that the number of conjugacy class equals the
number of irreducible representations of the finite group $G$, which indicates that the number of
pure single-loop excitation types (i.e., fluxes without attaching charges) equals the number of point-like excitations types.
By including the effect of charge, one can find the quantum dimension of a single-loop excitation
as follows:\cite{LanKongWen2017}
\be\label{d_xi}
d_{\chi;i}=|\chi|\cdot \mathrm{dim}[\mathrm{Rep}_i(G_{\chi})],
\ee
where $|\chi|$ denotes the order of the conjugacy class $\chi$.

(iii) Hopf-link loop excitations

For a Hopf-link loop excitation (see Fig.~\ref{Excitations}), there are two loops linked with each other,
with each loop carrying a flux labeled by the conjugacy class $\chi_{i=1,2}$.
These two conjugacy classes should commute with each other in the following sense\cite{JW1404,Wan1409}.
(See also Ref.~\onlinecite{Moradi1404} from the dimension reduction point of view.)
First, there exist $g\in\chi_1$ and $h\in \chi_2$ such that $gh=hg$.
Then one can find that the other elements $g':=kgk^{-1}\in \chi_1$ and $h':=khk^{-1}\in \chi_2, \, \forall k\in G$ also commute,
\textit{i.e.}, $g'h'=h'g'$.
To define the charge of the excitation,
first we define the centralizer with respect to $g$ and $h$ as $G_{g,h}:=\{x\in G| xg=gx,xh=hx\}$.
Then one can find that $G_{g,h}$ is isomorphic to $G_{g',h'}$, with $g',\,h'$ defined above.
It is convenient to denote these isomorphic centralizers as $G_{\chi_1,\chi_2}$. Then the charge carried by this Hopf-link loop excitation is described by the irreducible \textit{projective} representation of the centralizer $G_{\chi_1,\chi_2}$.

We note that for a non-Abelian gauge group $G$, there are some subtleties in characterizing a
Hopf-link loop excitation and defining its quantum dimension,
which is not yet well understood to our knowledge
\footnote{We thank Chenjie Wang for pointing out this issue to us.}.
In this work, for the Hopf-link loop excitations, we will mainly focus on the Abelian gauge group, \textit{e.g.},
$G=(\mathbb Z_N)^4$.
Then each group element is itself a conjugacy class,
and $G_{\chi_1,\chi_2}=G$.
Then the quantum dimension can be defined as\cite{JW1404,Wan1409,Jiang2014Generalized}
\be
d_{\chi_1,\chi_2;i}= \mathrm{dim}\left[\widetilde{\mathrm{Rep}}_{\chi_1,\chi_2;i}(G)\right].
\label{dxxiAbelian}
\ee
Here we use $\widetilde{\mathrm{Rep}}_i(G)$
to distinguish it from the linear representation $\mathrm{Rep}_i(G)$.
As will be shown later, the projective representation $\widetilde{\mathrm{Rep}}_i(G)$
carried by the Hopf-link loop excitation contains the information of the 4-cocycles $\omega$.
If the 4-cocycle $\omega$ in the DW theory is trivial, the projective
representation will automatically reduce to a linear representation
\footnote{As we will see later, even for nontrivial 4-cocycles, it is still possible that
the projective representation reduces to a linear representation, as long as the induced 2-cocycles are trivial.}.
Here we keep the index $\chi_1,\chi_2$ in \eqref{dxxiAbelian} to remind ourselves that the projective representation
depends on the conjugacy classes $\chi_1$ and $\chi_2$.
Remarkably, it has been understood that a higher dimensional $d_{\chi_1,\chi_2;i}$ in \eqref{dxxiAbelian} is closely
related with non-Abelian three-loop braiding in a DW theory\cite{JW1404}.

In this paper, through the entanglement entropy, we will detect the quantum dimensions in \eqref{d_i}, \eqref{d_xi}, and \eqref{dxxiAbelian} for different excitations.
There are more topologically complicated excitations for loop excitations. For example, the loop excitation can be a trefoil knot in Fig.~~\ref{Excitations}. Nevertheless, we restrict ourselves to only discussing the three simple cases listed above, since it is less clear to us how the complex knotted/linked loop excitations correspond to MESs on the $T^3$.

The rest of the paper is organized as follows. In Sec.~\ref{Sec: PartitionFunction}, we give a general consideration on the entanglement entropy in (3+1)D DW theory from the partition function point of view. In Sec.~\ref{Sec: Hamiltonian}, we give a brief introduction of the Hamiltonian version of (3+1)D DW theory and then introduce the form of MESs on a $T^3$. Then we study the effect of different excitations on the entanglement entropy in Sec.~\ref{Sec: 3dEE}. In particular, we study the (untwisted) DW theory with a non-Abelian Dihedral gauge group $G=D_3$, and the twisted type-IV DW theory with an Abelian gauge group $G=(\mathbb Z_N)^4$. Then we summarize and remark on the main results of this work in Sec.~\ref{Conclude}, and point out some future problems.

\section{Topological entanglement entropy in $(3+1)$D Dijkgraaf-Witten theories}
\label{Sec: PartitionFunction}

\begin{figure}[htp]
\includegraphics[width=3.40in]{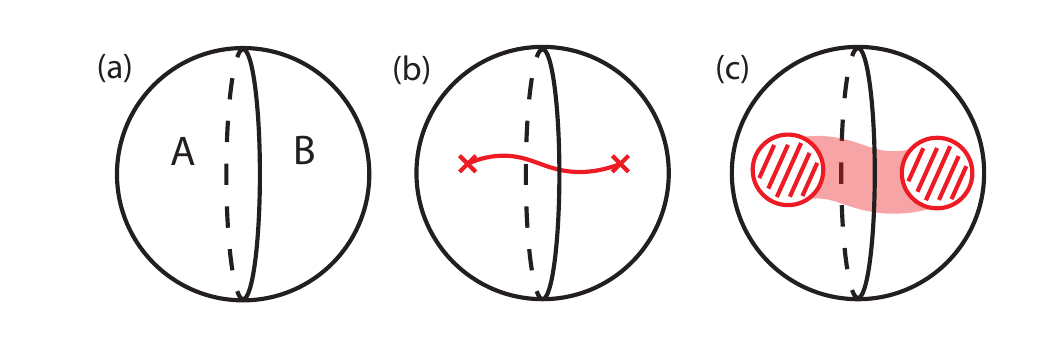}
\caption{ (Color online)
Bipartition of a spatial manifold $S^3$, where the interface between subsystems $A$ and $B$ is a $S^2$.
This spatial manifold $S^3$ can be viewed as the surface of a spacetime 4-manifold $D^4$, where the
wavefunctional is defined. The system may be in (a) a ground state, (b) an excited state with a pair of
point-like excitations, and (c) an excited state with a pair of loop excitations which may admit a
complicate knotted/linked structure. The two excitations are connected by a two dimensional membrane
with possible internal structure.
}
\label{S3}
\end{figure}

An elegant way to study the entanglement entropy in a (3+1)D DW theory is based on the surgery and the partition function method. The spirit is similar to the calculation of the entanglement entropy in a Chern-Simons theory in (2+1)D.\cite{Dong0802} It allows us to express the entanglement entropy in terms of partition functions of a (3+1)D DW theory on certain simple 4-manifolds.

Now we consider the ground state of a DW theory on a 3-sphere (in three spatial dimensions)
 described by the wavefunctional
$|\Psi\rangle$. By considering a bipartition into spatial subregions $A$ and $B$ (see Fig.~\ref{S3}), we may define the
reduced density matrix as $\rho_{A}=\text{tr}_{B}|\Psi\rangle \langle \Psi |$. Let the entanglement
cut between subregions $A$ and $B$ have the topology of a 2-sphere. By the general axioms of
TQFT\cite{Atiyah1988,Freed1992},
it is known that a path integral on $D^4$ produces the state $|\Psi\rangle$ in the Hilbert space
$\mathcal H_{S^3}$ which is 1-dimensional. We first notice that
$\text{tr}\,\rho_{A}=\text{tr}(\rho^n_{A})=\mathcal Z(S^4)$\cite{witten1989quantum,
Dong0802}. Then the $n$-th R\'enyi entropy is
\begin{align}
S^{(n)}_{A}:=\frac{1}{1-n}\ln\frac{\text{tr}\rho_A^n}{(\text{tr}\rho_A)^n}
=\ln \mathcal Z(S^4).
\label{TEE_sphere}
\end{align}
Since the result is independent of the R\'enyi index $n$, the entanglement entropy is the same as
the $n$-th R\'enyi entropy as
$
S_{A}^{\mathrm{topo}}=\lim_{n\to 1}S^{(n)}_{A}.
$
Here we use `$\mathrm{topo}$' in $S_{A}^{\mathrm{topo}}$ indicating that
the partition function method based on a TQFT only captures the TEE
for the ground state case, without including the area part of the entanglement entropy.

Let us specialize to the case of a DW theory $(G,\,\omega)$. Since the cohomology of the classifying space $BG$ is same as the group cohomology, we may write $\omega\in \mathcal{H}^{4}(BG,U(1))$. The classifying space has the property that isomorphism classes of flat $G$ bundles on a compact oriented manifold $M$ is equivalent to homotopy classes of maps $[\gamma]: M\to BG$.
Alternatively, $\gamma$ can be viewed as the homomorphism of the fundamental group $\pi_1(M)$ of the manifold
$M$ into the group $G$, up to conjugation.\cite{DW1990} 
Then the partition function for a DW theory can be 
defined as the sum over all possible $G$ bundles over $M$, weighted by $W=e^{2\pi i S}$ (here $S$ is the action of 
the DW theory)\cite{DW1990,wakui1992}
\begin{align}
\mathcal Z(M)
=\frac{1}{|G|}\sum_{[\gamma]} W(\gamma),
\end{align}
with
\begin{align}
W(\gamma)=\langle \gamma^{\star}\omega, [M] \rangle
\end{align}
where $\gamma^{\star}\omega \in \mathcal{H}^{4}(M,U(1))$ is the pullback of $\omega$ onto $M$ and $[M]$ denotes the fundamental homology class of $M$. In other words $\langle \gamma^{\star}\omega,[M]\rangle$ is a topological action that furnishes a pure $U(1)$ phase when evaluated on a specific choice of a flat bundle.  The normalization factor of $1/|G|$ can be justified by the fact that it gives the correct groundstate degeneracy for a three-sphere where $\text{GSD}(S^{3})=\mathcal Z(S^{3}\times S^{1})$. For $M=S^1\times S^3$, one has $\pi_1(S^1\times S^3)=\mathbb{Z}$. In addition, $\text{Hom}(\mathbb{Z},G)\simeq G$.
Then one has
\begin{equation}
Z(S^1\times S^3)=\frac{1}{|G|}\left(1\times |G|\right)=1.
\end{equation}
Here we have used $W(\gamma)=1$ i.e a cocycle evaluated on a trivial bundle is 1. \footnote{Firstly since $W(\gamma)$ is evaluated on a map in a particular homotopy class $[\gamma]: M\to BG$, it must by definition be insensitive to deformations that leave it in the same homotopy class. Then noticing that there are no non-trivial bundles on $S^3$, the map $\gamma$ can be deformed such that its pullback to $M$ gives a $G$-coloring of a triangulation of $M=S^3\times S^{1}$ which only has non-trivial elements of $G$ along $S^1$ and identity everywhere else. Now we recall that in general a $G$-coloring is an assignment of four elements of $G$ to each 4-simplex. Clearly all 4-simplices of the above coloring have atleast one (actually three) trivial elements. Furthermore since one builds Dijkgraaf-Witten theories from normalized cocycles i.e those such that for $[\omega]\in H_{\text{group}}(G,U(1))$, we require $\omega(1,g_1,g_2,g_3)=1$. Therefore $W(\gamma)=1$ for a $G$-bundle on $S^3\times S^{1}$.}

\bigskip\noindent For $G=(\mathbb Z_n)^k$ which is a main example we will study
in this work, partition functions take the explicit forms
\begin{align}\label{Z_234}
\mathcal Z_{\text{II,III}}(M)=&\;\frac{1}{|G|}\sum_{[A]}e^{\frac{2\pi ip}{n^2}\int_{M}A^{I}\cup A^J\cup \delta A^{K}},\quad p\in \mathbb{Z}_n\nonumber \\
\mathcal Z_{\text{IV}}(M)=&\; \frac{1}{|G|}\sum_{[A]}e^{\frac{2\pi ip'}{n}\int_M A^{I}\cup A^{J} \cup A^K \cup A^L}, \quad p'\in \mathbb{Z}_n.
\end{align}
Here the subscripts `II,III,IV' are meant to denote the partition functions corresponding to type-II,III,IV cocycles respectively\cite{JW1404,Propitius95}. $A^{I}\in \mathcal{H}^{1}(M,\mathbb Z_n)$ and $\delta A^I\in \mathcal{H}^{2}(M,n\mathbb Z)$ are obtained using a Bockstein homomorphism (For technical details of simplicial calculus and Bockstein homorphisms we refer the reader to appendix.A of Ref.~\onlinecite{kapustin2014bosonic})
\begin{align}
\beta: \mathcal{H}^{1}(M,\mathbb Z_n)\to \mathcal{H}^{2}(M,n\mathbb Z),
\end{align}
and $p,p'$ in Eq.~\eqref{Z_234} are a set of $\mathbb Z_n$ valued parameters that represent different choices
of cocycles $\omega\in \mathcal{H}^{4}(G,U(1))$. The integral may be evaluated by picking a triangulation of $M$ and
then explicitly assigning $U(1)$ phases to each 4-simplex via the cochain provided above. Further different
cochains are glued (i.e., multiplied) together and then we sum over all possible isomorphism classes of colorings of the triangulation. This amounts to summing over configurations $[A]\in \mathcal H^{1}(M,\mathbb Z_n)$.


In particular, $S^4$ admits a unique flat bundle which is trivial. Therefore,
\begin{align}\label{Z_S4}
\mathcal Z(S^4)=1/|G|.
\end{align}

Based on Eqs.\eqref{TEE_sphere} and \eqref{Z_S4},
the TEE for the ground state $|\Psi\rangle$ evaluates to
\begin{align}\label{TEE_sphere2}
S^{\mathrm{topo}}_A=-\ln |G|.
\end{align}
This generalizes previous results on an untwisted DW theory to a twisted DW theory.
It is noted that the entanglement cut here is chosen as a 2-sphere $S^2$. As shown in
Appendix \ref{Sec: S2_T2}, we also consider the type IV DW theory with $G=(\mathbb Z_N)^4$
with an entanglement cut $T^2$. It is found that the TEE
is the same as Eq.~(\ref{TEE_sphere2}).

Now we give several remarks of the entanglement entropies on the case with excitations.
First, the above method can be straightforwardly generalized when excitations are introduced.
Shown in Fig.~\ref{S3} (b) and (c) are examples of excited states with a pair of point-like excitations
and a pair of loop excitations with possible knotted or linked structure.
By repeating the above procedure, one can find the entanglement entropy
\be
S_A=\ln \mathcal{Z}(S^4, i),\quad i= \circ, \#,
\label{S_A01}
\ee
where $\circ$ represents a Wilson loop, and $\#$ represents a closed two dimensional Wilson membrane
that may potentially admit a complicate structure.
Since it is difficult to evaluate the partition function
$\mathcal{Z}(S^4, \circ)$ or $ \mathcal{Z}(S^4, \#)$ with Wilson loop/membrane inserted in
for a (3+1)D DW theory,
in order to investigate the effect of excitations on the quantum entanglement,
we resort to the exactly solvable model as discussed in the following sections.
Based on the calculation in Sec.~\ref{Sec: 3dEE}, the partition function $\mathcal{Z}(S^4,i)$ with $i=\circ,\#$
for a (3+1)D DW theory has the expression
$\mathcal{Z}(S^4,i)=d_i/|G|$,
where $d_i$ are the quantum dimensions of the topological excitations,
and then the entanglement entropy has the form:
\footnote{We emphasize that here the entanglement cut is a $S^2$, while the
entanglement cut considered in Sec.~\ref{Sec: 3dEE} is a $T^2$.
It is our future work to investigate if the effect of topological excitations on entanglement
entropy depends on the topology of entanglement cut or not, and if yes, how it depends on
the entanglement cut.
 }
$S_A=\ln d_i-\ln|G|$.

\section{(3+1)D Twisted Gauge Theory and MES}
\label{Sec: Hamiltonian}

The Hamiltonian version of a DW theory has been studied in both (2+1)D\cite{Hu2012,Mesaros1212}
and (3+1)D\cite{JW1404,Wan1409,Jiang2014Generalized}, where the canonical basis
for the degenerate ground states on $T^2$ (in (2+1)D) and $T^3$ (in (3+1)D) are also constructed.
By calculating the basis overlaps under the modular $\mathcal{S}$, $\mathcal{T}$ transformations,
one can obtain the  modular matrices of the underlying TQFT\cite{Hu2012,Wan1409} in the quasi-excitation basis.
For the purpose of our work, we will give a very brief review of this model by mainly focusing on the ingredients
of MESs on a $T^3$.
One can refer to Ref.~\onlinecite{Wan1409} for more details.

\begin{figure}[t]
\includegraphics[width=2.50in]{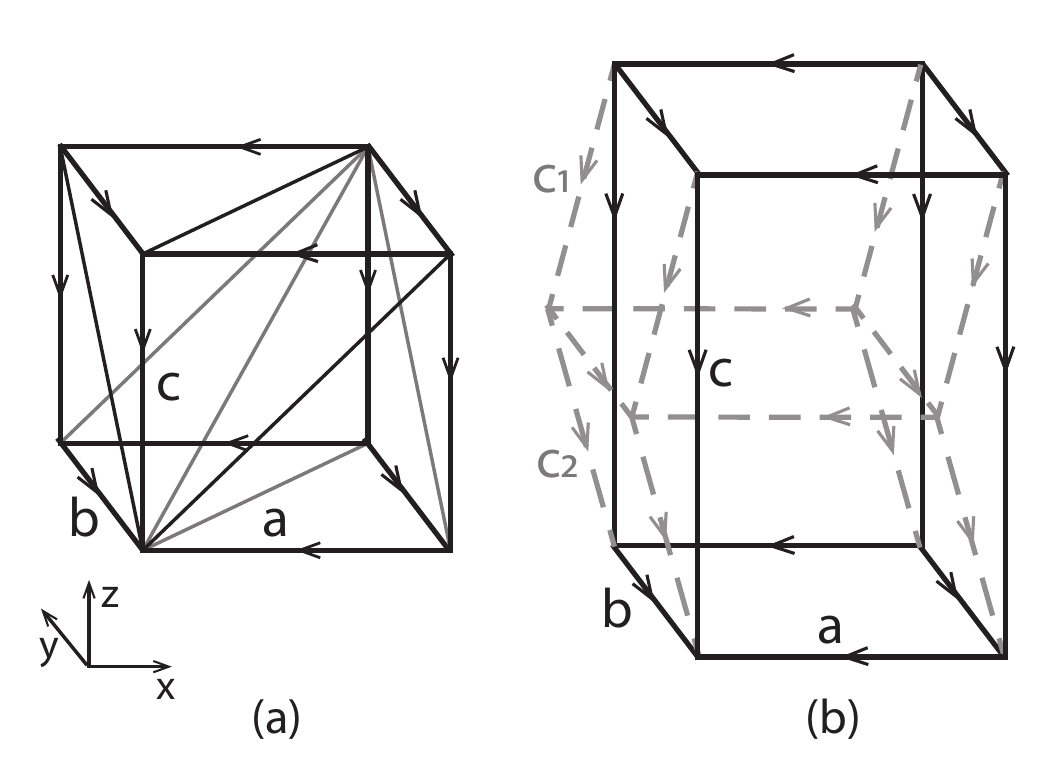}
\caption{
(a) Triangulation of a $T^3$ on a single cube with six tetrahedrons.
Periodic boundary conditions are imposed along each direction.
(b) Evolution from a single cube to two cubes, on which a three-torus is triangulated.
Not every bond in the triangulated $T^3$ is shown here.
}
\label{Triangulation}
\end{figure}

The model is characterized by $(H, G, \omega)$, where $H$ denotes the Hamiltonian, $G$ is a finite
(Abelian or non-Abelian) gauge group, and $\omega$ is a 4-cocycle.
When the 4-cocycle $\omega$ is trivial, the model reduces to an untwisted discrete gauge theory.
The Hilbert space is defined as follows:
\begin{equation}
\hilb = \otimes_b \hilb_b,	\quad
\hilb_b = \mathrm{Span}\lbrace \ket{g}| g\in G \rbrace,
\end{equation}
where $\hilb_b$ is the local Hilbert space on a bond $b$ of the triangulated lattice,
and the total Hilbert space is the tensor product of all local Hilbert spaces since we only consider bosonic systems in this work. In our convention, each basis $\ket{g}$ is associated with an orientation, and changing the orientation is equivalent to inversing the group element:
$
\ket{g,\rightarrow} = \ket{g^{-1},\leftarrow}.
$
This convention is same as in Refs.~\onlinecite{kitaev2003fault,Moradi1404,Wan1409}.
The group elements associated with all local Hilbert spaces can be interpreted as gauge fields in lattice gauge theory language.
The Hamiltonian is defined on a 3-dimensional triangulated manifold as follows\cite{Wan1409}
\begin{eqnarray}\label{eq.Hamiltonian}
H = - \sum_{v} A_v - \sum_{p} B_p,
\end{eqnarray}
where $A_v$ acts on the vertex $v$, and generates a gauge transformation
of the group element on each bond that connects to $v$,
 $B_p$ imposes the zero flux condition on each face (a triangle), and $[A_v,B_p]=0$.
The 4-cocycle $\omega$ is introduced when we act the operator
$A_v=\frac{1}{|G|}\sum_{[vv']=g\in G}A_v^g$  on vertex $v$,
by lifting the vertex $v$ to $v'$ with $[vv']=g\in G$.
The amplitude of this gauge transformation can be expressed
in terms of 4-cocycles $\omega$. (See Ref.\onlinecite{Wan1409} for details.)

Putting many details of this model aside, for our purpose, now we mainly focus on the MESs
on a triangulated $T^3$ (see Fig.~\ref{Triangulation}).
On the $T^3$, we have topologically protected degenerate ground states, due to the global degrees of freedom in the flat gauge field configurations. The MES, a canonical choice of ground state basis,  is labeled by three objects:\cite{JW1404,Wan1409,Jiang2014Generalized}
\begin{equation}\label{eq.quasiparticlebasis_untwisted}
\ket{\chi_1,\chi_2,\mu},
\end{equation}
where $\chi_1$ and $\chi_2$ denote the `flux' and $\mu$ denotes the `charge'.
Here $\chi_1$ and $\chi_2$ are two mutually commuting conjugacy classes of $G$,
and $\mu$ corresponds to the irreducible projective representation of the centralizer
$G_{\chi_1,\chi_2}$ as discussed in Sec.~\ref{Introduction}.
Now let us consider the MES in $z$-direction (see Fig.~\ref{Triangulation}).
First, it is noted that the conjugacy class $\chi_{1,2}$ can
be used to label the membrane, by considering that the loop excitation
is created by a membrane operator.
Then, $|\chi_1,\chi_2,\mu\rangle$ in $z$-direction may be viewed as
a state with membrane $\chi_1$ inserted in $yz$-plane, and membrane
$\chi_2$ inserted in $xz$-plane\cite{Jiang2014Generalized}.
Naively, one may consider that the `charge' $\mu$ corresponds to a Wilson loop inserted in $z$-direction.
However, this is not true in general, considering that this charge may be carried by a loop excitation,
and has no local source (or cannot be observed locally). The charge in this case is usually called
`Cheshire charge'\cite{Alford1990,Preskill199050,Preskill1992}.

The concrete form of $|\chi_1,\chi_2,\mu\rangle$ can be constructed as follows:\cite{JW1404,Wan1409,Jiang2014Generalized}
\be\label{MES3d}
\begin{split}
&|\chi_1,\chi_2,\mu\rangle
=\frac{1}{\sqrt{G}}\sum_{\substack{a\in \chi_1, b\in \chi_2;\\ c\in G_{a,b} }}
\mathrm{tr}\left[
\tilde{\rho}^{a,b}_{\mu}(c)
\right]
|a,b,c\rangle,
\end{split}
\ee
where $a$, $b$, and $c$ are the group elements living on the bonds in $x$, $y$, and $z$
directions, respectively (see Fig.~\ref{Triangulation} (a)).
$\tilde{\rho}^{a,b}_{\mu}$ is the irreducible projective representation of the centralizer $G_{a,b}$,
and is determined by
\be\label{PRtype4}
\tilde{\rho}^{a,b}_{\mu}(c)\tilde{\rho}^{a,b}_{\mu}(d)
=
\beta_{a,b}(c,d)
\tilde{\rho}^{a,b}_{\mu}(c\cdot d),
\ee
where $\beta_{a,b}(c,d)$, also called a factor-system, is a 2-cocycle obtained based on a slant product of the 4-cocycles $\omega$ (see Appendix \ref{Sec: 2cocycle}).
Now let us make a connection between $\tilde{\rho}^{a,b}_{\mu}(c)$ in Eq.~\eqref{MES3d} and
$\widetilde{\mathrm{Rep}}_{\chi_1,\chi_2;i}(G)$ for an Abelian group $G$
 in Eq.~\eqref{dxxiAbelian}. Since each group element is itself a conjugacy class, by considering $\chi_1=\{a\}$,
$\chi_2=\{b\}$, and $c\in G_{a,b}=G$, then $\tilde{\rho}^{a,b}_{\mu}(G)$ is exactly the same as
$\widetilde{\mathrm{Rep}}_{\chi_1,\chi_2;i}(G)$  in Eq.~\eqref{dxxiAbelian}.
For an untwisted gauge theory, $\tilde{\rho}^{a,b}_{\mu}$ reduces to the linear representation $\rho^{a,b}_{\mu}$. The basis in Eq.~\eqref{MES3d} gives the correct modular $\mathcal{S}$, $\mathcal{T}$ matrices when we compute the wave function overlaps under modular transformations\cite{Wan1409}.

Now let us discuss the MESs corresponding to the three types of excitations as mentioned in
Sec.~\ref{Introduction}, \textit{i.e.}, point-like excitations, single-loop excitations, and Hopf-link loop excitations.

\begin{figure}[t]
	\includegraphics[width=2.80in]{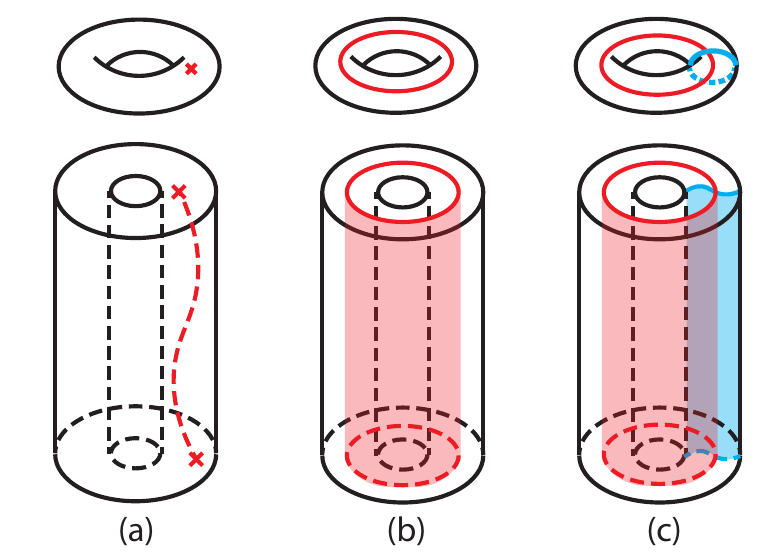}
	\caption{
		Lower panel: MESs on a three-torus corresponding to different types of excitations.
		Upper panel: By cutting the three-torus in $xy$-plane, one obtains excitations
		on the boundaries which are $T^2$, with
		(a) a point-like excitation, (b) a single-loop excitation, and (c) a Hopf-link loop excitation.
	}
	\label{MES}
\end{figure}

For a point-like excitation, both fluxes are trivial, \textit{i.e.}, $\chi_1=\chi_2=\{\mathbb 1\}$,
where $\mathbb 1$ represents the identity element in group $G$.
Then one has
\be\label{point_MES}
|\mathbb 1, \mathbb 1, \mu\rangle=\frac{1}{\sqrt{|G|}}\sum_{c\in G} \mathrm{tr}\left[\rho_{\mu}(c)\right]|\mathbb 1, \mathbb 1, c\rangle.
\ee
Here we use $\rho_{\mu}(c)$ instead of $\tilde{\rho}^{\mathbb{11}}_{\mu}(c)$, indicating that
$\tilde{\rho}^{\mathbb{11}}_{\mu}(c)$ reduces to the linear representation of $G$, as discussed in
Appendix \ref{Sec: 2cocycle}.
This case corresponds to a Wilson loop in representation $\mu$
threading through $z$-direction of the $T^3$, as shown in Fig.~\ref{MES}\,(a).
If we cut the $T^3$, one can find a point excitation on the boundary which is a $T^2$.

For a single-loop excitation, one of $\chi_1$ and $\chi_2$ is trivial while the other one is nontrivial. Without loss of generality, we choose
$\chi_1=\{\mathbb 1\}$. Then one has
\be
|\mathbb 1, \chi, \mu\rangle=\frac{1}{\sqrt{|G|}}\sum_{b\in \chi, c\in G_{\chi}} \mathrm{tr}\left[\rho^{\mathbb 1, b}_{\mu}(c)\right]|\mathbb 1, b, c\rangle.
\ee
Again, similar to the point-like excitations,
here the projective representation $\tilde{\rho}^{\mathbb 1, b}_{\mu}(c)$ reduces to a linear representation
$\rho^{\mathbb 1, b}_{\mu}(c)$ of $G_{\chi}$ (see Appendix \ref{Sec: 2cocycle}).
This case corresponds to the MES in Fig.~\ref{MES} (b).
A loop excitation will appear on the boundary $T^2$ when
we cut the $T^3$.

The third type of the excitations, \textit{i.e.}, a Hopf-link loop excitation, corresponds
to the MES in Eq.~\eqref{MES3d}, with $\chi_1\neq \{\mathbb 1\}$ and
$\chi_2\neq \{\mathbb 1\}$.
In this case, both fluxes are nontrivial. As shown in Fig.~\ref{MES}\,(c), when we cut the $T^3$,
two linked loop excitations which carry fluxes $\chi_1$ and $\chi_2$ respectively will appear on the boundary $T^2$.
It is noted that the MES with nontrivial $\chi_{1,2}$ in Eq.~\eqref{MES3d}
is characterized by the projective representation $\tilde{\rho}^{a,b}_{\mu}$, which is further determined by
the 2-cocycle $\beta_{a,b}(c)$ in Eq.~\eqref{PRtype4} and 4-cocycle through Eqs.\eqref{3cocycle} and \eqref{2cocycle}.
In other words, the information of 4-cocycle $\omega$ is contained in a Hopf-link loop excitation.
We will illustrate how this is reflected in the entanglement entropy in Sec.~\ref{Subsection: ZN4}.

\section{Effect of excitations on entanglement entropy}
\label{Sec: 3dEE}

\subsection{Untwisted DW gauge theory with $G=D_3$ }

Now we consider a discrete gauge theory with a dihedral group $D_3$,
which is a non-Abelian group and isomorphic to the symmetric group $S_3$.
Through this example, we aim to illustrate the effect of
(i) point-like excitations and (ii) single-loop excitations on the entanglement entropy.

The dihedral group $D_3$ is obtained by composing the six symmetries of an
equilateral triangle.
This group is generated by $\tau$ and $\theta$,
where $\tau$ is a flip about an axis passing through the center of the triangle and one of its vertices;  $\theta$ is
a rotation by $2\pi/3$ about the center of triangle.
There are in total six elements, which are generated as
$\mathbb 1$, $a=\theta$, $b=\theta^2$, $c=\tau$, $d=\tau\theta$, $e=\tau\theta^2$,
and three conjugacy classes: $\chi_{\mathbb 1}=\{\mathbb 1\}$, $\chi_2=\{c,d,e\}$
and $\chi_3=\{a,b\}$, with the corresponding centralizer subgroups
$G_{\chi_1}=D_3$, $G_{\chi_2}=\mathbb Z_2$, and $G_{\chi_3}=\mathbb Z_3$.

As studied in Ref.\onlinecite{LanKongWen2017}, there are three types of point-like excitations (denoted by $p_0$,
$p_1$, and $p_2$) and five types of single-loop excitations (denoted by $s_{20}$, $s_{21}$, $s_{30}$, $s_{31}$, and $s_{32}$).
For the point-like excitations, $p_0$ corresponds to the trivial (one dimensional) representation of $D_3$,
$p_1$ and $p_2$ correspond to the one-dimensional and two-dimensional representations of $D_3$.
The loop excitations are labeled by $(\chi,\mathrm{Rep}_i(G_{\chi}))$ as follows:
$s_{20}=(\chi_2,\mathrm{Rep}_0(\mathbb Z_2))$,
$s_{21}=(\chi_2,\mathrm{Rep}_1(\mathbb Z_2))$,
$s_{30}=(\chi_3,\mathrm{Rep}_0(\mathbb Z_3))$,
$s_{31}=(\chi_3,\mathrm{Rep}_1(\mathbb Z_3))$,
and $s_{32}=(\chi_3,\mathrm{Rep}_2(\mathbb Z_3))$.
The quantum dimensions of these point-like and loop excitations are, according to Eqs.\eqref{d_i} and \eqref{d_xi},
\be\label{d_D3}
\begin{split}
&d_{p_0}=d_{p_1}=1, \quad d_{p_2}=2,\\
&d_{s_{20}}=d_{s_{21}}=3,\\
&d_{s_{30}}=d_{s_{31}}=d_{s_{32}}=2.
\end{split}
\ee
It is interesting that each loop excitation can shrink to point-like excitations.\cite{LanKongWen2017}
For example, the loop excitation $s_{21}$ can shrink to point-like excitations:
$s_{21}\rightarrow p_1\oplus p_2$, during which the quantum dimension is conserved by considering that
$d_{s_{21}}=d_{p_1}+d_{p_2}$. However, it is noted that in the configuration in Fig.~\ref{MES},
the loop excitation on the boundary, which is a $T^2$, is along the non-contractible circle, and therefore it
cannot shrink to point-like excitations.

Here we give a sample calculation on the entanglement entropy for an MES
corresponding to the point-like excitation $p_2$,
the quantum dimension of which is $d_{p_2}=2$.
Other cases can be straightforwardly calculated in the same way.
To write down the MES corresponding to $p_2$,
we first need to check the nonzero characters $\mathrm{tr}\left[\rho_{\theta}(g)\right]$, $\forall g\in D_3$,
where we denote the 2-dimensional representation of $D_3$ as $\rho_\theta$. It can be found
that the non-zero characters are:
\be
\mathrm{tr}\left[\rho_{\theta}(\mathbb 1)\right]=2, \quad \mathrm{tr}\left[\rho_{\theta}(a)\right]=\mathrm{tr}\left[\rho_{\theta}(b)\right]=-1.
\ee
Then based on Eq.~\eqref{point_MES}, the MES on a single-cube $T^3$ has the expression
\be
|\mathbb 1,\mathbb 1,\theta\rangle=\frac{1}{\sqrt{|G|}}|\mathbb 1\rangle\otimes
|\mathbb 1\rangle \otimes\Big(2|\mathbb 1\rangle-|a\rangle-|b\rangle\Big),
\ee
where for convenience we have written $|\mathbb 1,\mathbb 1,c\rangle$ as
$|\mathbb1\rangle\otimes |\mathbb 1\rangle\otimes |c\rangle$,
with $|\mathbb 1\rangle\otimes |\mathbb 1\rangle$ representing the trivial fluxes in both $x$ and $y$ directions.
The next step is to write this MES in a two-cube basis, so that we can study the entanglement between two
cubes. This can be done by writing
\be\label{1to2_untwist}
|c\rangle\to \frac{1}{\sqrt{|G|}}\sum_{\substack{c_1,c_2\in G;\\   c_1\cdot c_2=c}}|c_1\rangle\otimes |c_2\rangle,
\ee
where $c$, $c_1$ and $c_2$ are group elements defined on the bond in $z$-direction, as shown in Fig.~\ref{Triangulation}\,(b).
In this two-cube basis, one can obtain the reduced density matrix for cube $A$ as $\rho_A=\mathrm{tr}_{\bar{A}}
\big(|\mathbb 1,\mathbb 1,\theta\rangle\langle \mathbb 1,\mathbb 1, \theta|\big)$.
One can find the eigenvalues of $\rho_A$ as $0$ with multiplicity two, and $1/4$ with multiplicity four.
Then the entanglement entropy for cube $A$ is $S_A=\frac{1}{1-n}\ln \left(\mathrm{tr}\rho_A^n\right) =2\ln 2$.
By repeating this procedure for all the other point-like and single-loop excitations, one can find
\be\label{SA_D3}
S_A=2\ln d_i,
\ee
where $d_i$ are the quantum dimensions of the corresponding excitations, as given in Eq.~\eqref{d_D3}.

There are two remarks on the result in Eq.~\eqref{SA_D3}: (i) The factor `2' in $S_A$
arises from the fact that there are two interfaces when bipartitioning a torus. In other words, the Wilson loop/membrane operators across the entanglement cut are detected twice.
(ii) One may ask why both the area-law term and the TEE term are not included in Eq.~\eqref{SA_D3}. The reason is that these two terms cancel with each other by considering that $2V\ln |G|-2\ln |G|=0$ when $V=1$ (see also Sec.~\ref{Introduction} above Eq.~\eqref{Gamma}). As a simple exercise, one can check the MESs on cubes with $V=2$, and there is an extra piece $2V\ln |G|-2\ln |G|=2\ln |G|$ in Eq.~\eqref{SA_D3}. Putting the area part of the entanglement entropy back to Eq.~(\ref{SA_D3}) leads to the general formula of the entanglement entropy:
\begin{equation}
S_A= 2V \ln |G| + 2\ln d_i - 2\ln|G|.
\end{equation}

\subsection{Twisted DW theory with $G=(\mathbb Z_N)^4$}
\label{Subsection: ZN4}

For a (3+1)D twisted DW gauge theory with an Abelian gauge group $G=(\mathbb Z_N)^4$, naively, one may expect that all the excitations are Abelian with quantum dimensions $1$. What is interesting, in Ref.~\onlinecite{JW1404}, it was found that when specific 4-cocycles $\omega$ are introduced (the so-called type IV 4-cocycle), there exist loop excitations with a quantum dimension $N$ ($N$ is prime), which is related with the non-Abelian three-loop braiding. It turns out these excitations correspond to the Hopf-link loop excitations in Fig.~\ref{MES} (c).

Since $G=(\mathbb Z_N)^4$ is Abelian,
each group element is itself a conjugacy class, and then the MES in Eq.~\eqref{MES3d} can be simplified as
\be\label{MES3d_Abelian}
\begin{split}
|a,b,\mu\rangle=\frac{1}{\sqrt{|G|}}\sum_{c \in G}
\mathrm{tr}\left[\tilde{\rho}_{\mu}^{a,b}(c)\right]|a,b,c\rangle,
\end{split}
\ee
where $a,b\in G$ are used to label the flux in $x,y$ directions, and $\mu$ is used to label the charge
representation.
In this theory, there are several kinds of
4-cocycles $\omega\in \mathcal{H}^{4}(G;\,U(1))$, which are named type II, III, and IV
4-cocycles in Ref.\onlinecite{JW1404}.
Then the 2-cocycles $\beta_{a,b}(c,d)$ that enter the definition of
$\tilde{\rho}^{a,b}_{\mu}$ in Eq.~\eqref{PRtype4} can be obtained by doing slant
product twice from 4-cocycles $\omega$ (See Appendix \ref{Sec: 2cocycle} for more details.).

It was found that only the type IV 4-cocycles may lead to a higher dimensional
irreducible projective representation $\tilde{\rho}^{a,b}_{\mu}$.
To be concrete, the type IV 4-cocycle has the form
\be
\omega_{\text{IV}}(a,b,c,d)=\exp\left(\frac{2\pi i p }{N}a_1b_2c_3d_4\right).
\ee
For simplicity, let us choose the nontrivial parameter
$p=1$ from now on. A general $p\in \mathbb Z_N$ can be discussed in a similar way.
From a slant product (see Appendix \ref{Sec: 2cocycle}), one can find the induced 2-cocycles as follows
\be\label{c_ab(cd)}
\begin{split}
&\beta_{a,b}(c,d)\\
=&\exp\Big\{\frac{2\pi i}{N}\big[
(a_4b_3-a_3b_4)c_1d_2+(a_2b_4-a_4b_2)c_1d_3\\
&+(a_4b_1-a_1b_4)c_2d_3+(a_3b_2-a_2b_3)c_1d_4\\
&+(a_1b_3-a_3b_1)c_2d_4+(a_2b_1-a_1b_2)c_3d_4
\big]
\Big\}.
\end{split}
\ee
From Eq.~\eqref{PRtype4}, each 2-cocycle $\beta_{a,b}$ specifies a class of irreducible
projective representations for $G$.
As we will see later, depending on the choice of the flux $a$ and $b$,
the dimension of the irreducible projective representation $\tilde{\rho}^{a,b}_{\mu}(c)$
is $1$ or $N$.

\subsubsection{Excitations with quantum dimensions 1}

Based on Eq.~(\ref{c_ab(cd)}), one can find that for an arbitrary $c$ and $d$,
$\beta_{a,b}(c,d)$ will become trivial, \textit{i.e.}, $\beta_{a,b}(c,d)=1$ ,
when satisfying the following condition
\be\label{TrivialCondition}
a_ib_j=a_jb_i,\quad \forall i\neq j.
\ee
For these cases, the irreducible projective representations $\tilde{\rho}^{a,b}_{\mu}(c)$
become linear representations of $G=(\mathbb Z_N)^4$ that are one dimensional.

Now let us check the quantum dimensions for point-like and single-loop excitations in this theory.
For a point-like excitation, one has $a=b=\mathbb 1$, where $\mathbb 1:=(0000)$ denotes the identity
group element in $(\mathbb Z_N)^4$.
For a single-loop excitation, either $a=\mathbb 1$ or $b=\mathbb 1$.
In both cases, Eq.~\eqref{TrivialCondition} is satisfied.
As a result, $\beta_{a,b}(c,d)$ is trivial for both point-like and single-loop excitations,
and their quantum dimensions are 1.

The entanglement property for an MES with one dimensional quantum dimension
 has been discussed in Ref.\onlinecite{Jiang2014Generalized}.
For the completeness, we give a brief review here.
For one dimensional irreducible projective representation,
one has $\mathrm{tr}\left[\tilde{\rho}_{\mu}^{a,b}(c)
\right]=\tilde{\rho}_{\mu}^{a,b}(c)$.
Then Eq.~(\ref{PRtype4}) can be rewritten as
\be\label{PRtype4_ch}
\mathrm{tr}\left[\tilde{\rho}^{a,b}_{\mu}(c)\right]\cdot \mathrm{tr}\left[\tilde{\rho}^{a,b}_{\mu}(d)\right]
=
\beta_{a,b}(c,d)
\cdot
\mathrm{tr}\left[\tilde{\rho}^{a,b}_{\mu}(c\cdot d)\right].
\ee
By triangulating the $T^3$ with two cubes in Fig.~\ref{Triangulation} (b) ,
each one-cube basis $|a,b,c\rangle$ in Eq.~\eqref{MES3d_Abelian} becomes\cite{Jiang2014Generalized}
\be\label{1to2}
|a,b,c\rangle\to\frac{1}{\sqrt{|G|}}\sum_{\substack{c_1,c_2\in G;\\ c_1\cdot c_2=c}}\beta_{a,b}(c_1,c_2)|a,b,c_1,c_2\rangle.
\ee
Note that compared to the case of untwisted gauge theory in Eq.~\eqref{1to2_untwist}, there is an extra $U(1)$
phase $\beta_{a,b}(c_1,c_2)$ for the twisted case.
Here $\beta_{a,b}(c,d)$ is the same as the 2-cocycle that appears in Eq.~\eqref{PRtype4}, which can be understood
as the phase associated with the triangulated manifold in Fig.~\ref{Triangulation}\,(b) after an appropriate
coloring of the manifold has been made.\cite{Jiang2014Generalized}

Then the MES in Eq.~(\ref{MES3d_Abelian}) can be rewritten as
\be
\begin{split}
|a,b,\mu\rangle=&\frac{1}{|G|}
\sum_{c_1,c_2 \in G}
\mathrm{tr}\left[\tilde{\rho}_{\mu}^{a,b}(c_1\cdot c_2)\right]\beta_{a,b}(c_1,c_2)|a,b,c_1,c_2\rangle\\
=&\frac{1}{|G|}
\sum_{c_1,c_2 \in G}
\mathrm{tr}\left[\tilde{\rho}_{\mu}^{a,b}(c_1)\right]
\mathrm{tr}\left[\tilde{\rho}_{\mu}^{a,b}(c_2)\right]
|a,b,c_1,c_2\rangle,
\end{split}
\ee
where we have used Eq.~\eqref{PRtype4_ch}.
It is found that $|a,b,\mu\rangle$ is a direct product state for the two cubes, and the entanglement entropy for
a single cube $A$ is $S_A=0$, \textit{i.e.}, one cannot detect the point-like and single-loop excitations
because of their trivial quantum dimensions.
Again, as remarked below Eq.~\eqref{SA_D3}, here the area-law term cancels with the
TEE term.

\subsubsection{Hopf-link loop excitations with higher quantum dimensions}

For Hopf-link loop excitations, both $a$ and $b$ correspond to nontrivial fluxes (see Fig.~\ref{MES}).
Then it is possible to have nontrivial 2-cocycles $\beta_{a,b}(c,d)$, which may result in higher quantum dimensions
$d_{\chi_1,\chi_2; i}$ in Eq.~\eqref{dxxiAbelian}.
\footnote{It is noted that not all Hopf-link loop excitations have higher quantum dimensions.
For example, for nonvanishing flux $a=b$, Eq.~\eqref{TrivialCondition} still holds. Then
the 2-cocycle $\beta_{a,b}$ is trivial, and one still has one dimensional quantum dimension.
}

For $G=(\mathbb Z_N)^4$, we choose the nonvanishing flux
$a=(0100)$, and $b=(1000)$ to have a non-trivial 2-cocycle
\be\label{c_abcd}
\beta_{a,b}(c,d)=\exp\left(\frac{2\pi i}{N} c_3d_4\right).
\ee
Based on Eq.~\eqref{PRtype4}, it can be easily checked that
\be
\tilde{\rho}^{a,b}(ij10)\cdot \tilde{\rho}^{a,b}(ij01)=\omega
\tilde{\rho}^{a,b}(ij01)\cdot \tilde{\rho}^{a,b}(ij10),
\ee
where $i,j\in \mathbb Z_N$, and we have defined $\omega:=\exp\left(\frac{2\pi i}{N}\right)$.
That is, $\tilde{\rho}^{a,b}(ij01)$ and $\tilde{\rho}^{a,b}(ij10)$
do not commute with each other, which indicates that the projective representation
$\tilde{\rho}^{a,b}$ are necessarily higher dimensional.
By checking all the $\tilde{\rho}^{a,b}(c)$ that satisfy
Eq.~\eqref{PRtype4}, one can find there are $N^2$ inequivalent
irreducible projective representations labeled by $(p,q)$ where
$p,q\in \mathbb Z_N$. In the representation $(p,q)$, one has
$\tilde{\rho}^{a,b}_{(p,q)}(0100)=\omega^p \mathbb{I}_N$, and
$\tilde{\rho}^{a,b}_{(p,q)}(1000)=\omega^q \mathbb I_N$.
where $\mathbb I_N$ is a $N$ dimensional identity matrix.
In addition, all the $\tilde{\rho}^{ab}_{(p,q)}(c)$ that have non-zero characters are of the form
$\tilde{\rho}^{a,b}_{(p,q)}(ij00)=\omega^{j\cdot p+i\cdot q}\mathbb I_N$, $\forall i,j\in \mathbb Z_N$.
For other $\tilde{\rho}^{a,b}_{(p,q)}(ijmn)$ with $m\neq 0$ or $n\neq 0$, the
characters will be zero.
Here, for simplicity, let us check the representation with $(p,q)=(0,0)$.
Then the corresponding MES has a simple form
\be\label{MES_typeIV}
|a,b,\mu\rangle=\frac{N}{\sqrt{|G|}}\sum_{i,j=0}^{N-1}|a,b,ij00\rangle,
\ee
where we have used the fact that $\mathrm{tr}[\tilde{\rho}^{a,b}_{(0,0)}(ij00)]=N$, and here $|G|=N^4$.
Based on Eq.~\eqref{1to2},
we may rewrite the MES in terms of two-cube basis as follows
\be
\begin{split}
|a,b,\mu\rangle
&=\frac{N}{\sqrt{|G|}}\cdot \frac{1}{\sqrt{|G|}}
\underbrace{|a\rangle\otimes |b\rangle}_{\text{flux}}\otimes\sum_{k,l,m,n=0}^{N-1}\sum_{i,j=0}^{N-1}
\omega^{m\cdot [N-n]}\\
&\underbrace{|klmn\rangle}_{\text{cube }\bar{A}}\otimes
\big|\underbrace{[N-k+i][N-l+j][N-m][N-n]}_{\text{cube A}}\big\rangle,\nonumber
\end{split}
\ee
where $[i]:=i~\mathrm{mod}~ N$, and the term $\omega^{m\cdot [N-n]}$ comes from the 2-cocycle
in Eq.~\eqref{c_abcd}.
Then after some simple algebra, one can find that
the reduced density matrix for cube $A$ may be written as
$\rho_{A}=\bigoplus_{m,n=0}^{N-1}\rho_{mn}$,
where
\be
\begin{split}
\rho_{mn}=&\frac{1}{N^4}\underbrace{|a,b\rangle\langle a,b|}_{\text{flux}}\otimes\Big(
\sum_{k,l=0}^{N-1}\big |kl[N-m][N-n]\big\rangle
\Big)\\
&\Big(
\sum_{u,v=0}^{N-1}\big\langle uv[N-m][N-n]\big|
\Big).
\end{split}\nonumber
\ee
Each $\rho_{mn}$ is a $N^2$ dimensional matrix with all elements being $1/N^4$. The eigenvalues of each $\rho_{mn}$ are
$
\lambda=\underbrace{0,0,\cdots, 0}_{N^2-1}, \frac{1}{N^2}.
$
Therefore, the eigenvalues of $\rho_{\text{A}}$ are
\be
\lambda=\underbrace{0,\cdots, 0}_{(N^4-N^2)}, \underbrace{\frac{1}{N^2},\cdots,\frac{1}{N^2}}_{N^2 }.
\ee
Then the $n$-th R\'enyi entropy for cube $A$ (or equivalently cube $\bar{A}$) has the form
\be\label{SA_Hopf}
S_{\text{A}}^{(n)}=\frac{1}{1-n}\ln \text{Tr}\left(\rho_{\text{A}}^n\right)
=2\ln N.
\ee
Again, the factor `2' appears because there are two interfaces for a bipartite $T^3$.
As before, here the area-law term cancels with the TEE term [see
also the remarks below Eq.~\eqref{SA_D3})].
The result in Eq.~\eqref{SA_Hopf} agrees with the observation in Ref.\onlinecite{JW1404}
that the  non-Abelian string-like excitations with nontrivial 2-cocycle $\beta_{a,b}$ have quantum dimensions $N$.
In Ref.\onlinecite{JW1404}, the quantum dimensions $d_i=N$ are obtained based on
the ground state counting on a three-torus $T^3$.
Here, we detect this quantum dimension based on the entanglement entropy.

The calculation above can be straightforwardly generalized to
 other irreducible projective representations $\mu=(p,q)$ with $\tilde{\rho}^{a,b}_{\mu}(0100)=\omega^p \mathbb I_N$
and $\tilde{\rho}^{a,b}_{\mu}(1000)=\omega^q \mathbb I_N$.
One needs to introduce a $U(1)$ phase $\omega^{j\cdot p+i\cdot q}$ in front of the basis
$|a,b,ij00\rangle$ in Eq.~\eqref{MES_typeIV}.
 It can be found that the entanglement entropy $S_A$
 is the same as Eq.~\eqref{SA_Hopf}.

As a short summary for the case of type IV twisted gauge theory with $G=(\mathbb Z_N)^4$, both the point-like and single-loop excitations have quantum dimension $1$, and have no contribution to the entanglement entropy. For Hopf-link excitations with nonvanishing fluxes $a$ and $b$, when the 2-cocycle $\beta_{a,b}$ in Eq.~\eqref{c_ab(cd)} are nontrivial, these excitations will have quantum dimensions $N$, and contribute $\ln N$ to the entanglement entropy.

It is noted that for other types of 4-cocycles, \textit{e.g.}, type II and type III 4-cocycles, since all the corresponding  2-cocycles $\beta_{a,b}$ are trivial, then
the Hopf-link loop excitations will have quantum dimensions 1,\cite{JW1404} and have no contribution to the entanglement entropy. From this point of view, the entanglement entropy with a Hopf-link loop excitation can be used to distinguish certain type IV 4-cocycles from the others. To obtain further details of the 4-cocycles from the entanglement entropy, one may need to study loop excitations with more interesting structures.

\section{Concluding remarks}
\label{Conclude}

In summary, we study the entanglement entropy of (3+1)D topological orders in this paper which is realized in (3+1)D DW theories labeled by $(G,\,\omega)$. Here $G$ is a finite gauge group and $\omega\in\mathcal{H}^4[G;U(1)]$ is a group 4-cocycle. We study the entanglement entropy with and without excitations. Depending on the types of excitations, it is found that the entanglement entropy can capture different information of $(G,\,\omega)$ as follows:

\begin{enumerate}

\item
For a generic DW theory on a spatial manifold $S^3$, we find that the TEE in the ground state is $\gamma=-\ln |G|$, regardless of the 4-cocycle $\omega$. In other words, the ground state entanglement entropy captures the information of
group order $|G|$.

\item
For a point-like excitation, it contributes $\ln d_i$ to the entanglement entropy, where $d_i=\mathrm{dim}\left[\mathrm{Rep}_i(G)\right]$. In other words, the entanglement entropy with point-like excitations captures the dimension of irreducible representation of $G$, \textit{i.e.}, $|\mathrm{Rep}(G)|$.

\item
For a single-loop excitation, its contribution to the entanglement entropy is $\ln d_{\chi,i}$ with $d_{\chi,i}=|\chi|\cdot \mathrm{dim}[\mathrm{Rep}_i(G_{\chi})]$. In other words, the entanglement entropy with single-loop excitations capture the information of conjugacy class size $|\chi|$ and the dimension of irreducible representation of its centralizer $G_{\chi}$.

\item
For a Hopf-link loop excitation, we mainly focus on the case with Abelian gauge group $G$. Its contribution to the entanglement entropy is $\ln d_{\chi_1,\chi_2;i}$, where
$d_{\chi_1,\chi_2;i}= \mathrm{dim}\left[\widetilde{\mathrm{Rep}}_{\chi_1,\chi_2;i}(G)\right]$. That is, the dimension of irreducible \textit{projective} representations of $G$ can be detected. This dimension depends on the 4-cocycles $\omega$ through Eq.~\eqref{PRtype4}. In other words, the entanglement entropy with Hopf-link loop excitations captures the information of 4-cocycles $\omega$.

\end{enumerate}

Now we give some remarks on the above results:

First, the entanglement entropy with point-like and single-loop excitations can capture different information of the finite gauge group $G$. However, they cannot capture the information of 4-cocyles $\omega$. The underlying reason is that the charges of these two excitations only correspond to \textit{linear} representation of $G$ or $G_{\chi}$. They are not `twisted' by the 4-cocycles $\omega$. On the other hand, for the Hopf-link loop excitations, they carry irreducible \textit{projective} representations of $G$, which are `twisted' by the 4-cocycles $\omega$. This is why we can detect information of $\omega$ through entanglement entropy of Hopf-link loop excitations.

Second, we would like to emphasize that the Hopf-link loop excitations cannot capture \textit{all} the information of 4-cocycles $\omega$. The reason is simple: One can find that the projective representations carried by Hopf-link loop excitations are directly determined by the 2-cocycles $\beta_{a,b}(c,d)$ in Eq.~\eqref{PRtype4}. During the dimension reduction from 4-cocycles $\omega$ to 2-cocycles $\beta$, not \textit{all} the information of $\omega$ is kept. This is why we can only distinguish \textit{certain} type-IV 4-cocycles $\omega$ from the others in the case of $G=(\mathbb Z_N)^4$. To detect more information of the 4-cocycles, we expect that the excitations with more interesting (and complicate) structures should be considered.

Third, it is natural to ask what happens if the theory under consideration deviates from the DW theory at the fixed point.
In this case, when the entanglement cut is a two-torus $T^2$, 
the constant correction to the area-law term 
in entanglement entropy includes the TEE as well as the contribution from curvature of the 
entanglement cut, \textit{etc}.
The way to extract the TEE in this case has been discussed in 
Refs.\onlinecite{Grover1108,Zheng2017Structure}.
It was shown that the topological contribution can be extracted via a 
generalized Kitaev-Preskill-Levin-Wen approach.\cite{levin2006,kitaev2006}
Moreover, it has been shown that when the theory flows to the infrared fixed point, the mean curvature contribution vanishes, and the constant part is the TEE.\cite{Zheng2017Structure}

\bigskip
There are many interesting problems we hope to solve in the future, and we mention
some of them as follows.

\begin{enumerate}
\item It is interesting to study various choices of entanglement cuts.
In this work, all excitations lie inside the subsystems.
It is natural to ask what happens if the entanglement cut is across the loop excitation itself.
One also needs to understand how the effect of excitations on entanglement entropy depends on
the topology of the entanglement cut.

\item The effect of excitations on the entanglement entropy is studied based on MES in this work.
 It would be very interesting to explore the partition function method based on a TQFT, in which
one needs to evaluate the partition functions on various 4-manifolds
with Wilson loops/membranes inserted.

\item It is desirable to classify or characterize various loop excitations with interesting knotted and linked structure
in (3+1)D topological orders.
For example, how are different excitations labeled by data based on $(G,\,\omega)$?
A good understanding of these excitations may help to detect more
information of the underlying theory through the entanglement entropy.
\end{enumerate}

\section{Acknowledgement}

We thank Tian Lan, Shinsei Ryu, Senthil Todadri,
and Xiao-Gang Wen for helpful discussions on related topics.
In particular, we thank Chenjie Wang for discussions on the properties of various loop excitations in (3+1)D
topological orders, and Juven Wang for discussions and sharing notes on the partition
function of (3+1)D BF theories, and bringing our attention to the upcoming work.\cite{WangJ1801}
We thank the workshop on Strongly Correlated Topological Phases of Matter at Simons center,
where part of this work was carried out.
X.W. was supported by the postdoc fellowship from
Gordon and Betty Moore Foundation EPiQS Initiative through Grant No.~GBMF4303 at MIT.
P.Y. was supported in part by the National Science Foundation grant DMR 1408713 and DMR 1725401 at the University of Illinois, and grant of the Gordon and Betty Moore Foundation EPiQS Initiative through Grant No. GBMF4305.

\appendix

\section{On the TEE in the ground state
of Type IV DW theory}
\label{Sec: S2_T2}

In this appendix, we consider the entanglement entropy in the ground state of type IV Dijkgraaf-Witten model. We argue that the topological part of the entropy does not depend on the genus of the entanglement surface. In particular, $S_{\mathrm{topo}}[S^2]=S_{\mathrm{topo}}[T^2]$, where $S^2$ and $T^2$ are entanglement surfaces.

The type IV DW model with gauge group $\otimes_{I=1}^4\mathbb{Z}_{n_I}$ is described by the following Lagrangian,
\be
\begin{split}
\lag&= \sum_{I=1}^4 \frac{n_I}{2\pi}B_IdA_I+ \frac{n_1n_2n_3n_4 p}{(2\pi)^3 \gcd(n_1, n_2, n_3, n_4)}A_1A_2A_3A_4,\\
&=:\mathcal{L}_0+\mathcal{L}_1,\nonumber
\end{split}
\ee
where $p\in \mathbb{Z}_{\gcd(n_1, n_2, n_3, n_4)}$. We consider the spacetime to be $D^4$, whose boundary is $\partial D^4=S^3$ which we identify as the spatial slice. Then the ground state on $S^3$ is
\begin{eqnarray}
|\psi\rangle=\mathfrak{C}\sum_{\mathcal{C}, \mathcal{C}'}\int_{\mathcal{C'}|_{S^3}}\mathcal{D}B_I\int_{\mathcal{C}|_{S^3}}\mathcal{D}A_I\exp\bigg(i
\int_{D^4}\mathcal{L}d^4x
\bigg)|\mathcal{C}\rangle,
\nonumber
\end{eqnarray}
where $\mathfrak{C}$ is the normalization factor. $\mathcal{C}$ and $\mathcal{C}'$ are the configurations for the $A$-cochain and $B$-cochain respectively. The path integral of $A$-cochain is over all the configurations of $A$ over $D^4$ with fixed configurations on the $S^3$, and similar for $B$-cochain.
Integrating over $B_I$, we get
\begin{equation}\label{WF01}
|\psi\rangle=\mathfrak{C}\sum_{\mathcal{C}}\int_{\mathcal{C}|_{S^3}}\mathcal{D}A_I\delta(dA_I)\exp\bigg( i \int_{D^4} \mathcal{L}_1d^4x\bigg)|\mathcal{C}\rangle\quad
\end{equation}
The delta function in the path integral implies that the configuration of $A$ field is a sum of closed three volumes in the dual lattice if it does not intersect with the spatial slice. On the spatial slice, the configuration of $A$ field is a sum of closed two-surfaces in the dual lattice.
\begin{eqnarray}
A_I=\frac{2\pi}{n_I} \sum_{i}*_4\Sigma(V^I_i)=\frac{2\pi}{n_I} \sum_{i}*_3\Sigma(S^I_i),
\end{eqnarray}
where $\partial V^I_i=S^I_i$, and the summation of $i$ runs over all possible volume (surface)
configuration in the dual lattice. $\Sigma(V_i^I)$ is a delta function which is one only if it is evaluated on the dual-lattice three volume $V_i^I$. $*_4$ is the Hodge dual operator in four dimensions, hence $*_4\Sigma(V^I_i)$ is a one form.
By plugging $A_I$ into the ground state in Eq.~\eqref{WF01}, one can find that
the wavefunction $|\psi\rangle$ is the sum of different configurations weighted by ``intersection numbers"
of three closed dual lattice surfaces and one volume bounded by a dual-lattice surface in $S^3$:
\begin{widetext}
\begin{eqnarray}\label{WFof IVDW}
|\psi\rangle=\mathfrak{C}\sum_{\mathcal{C}}\exp\bigg( i \frac{2\pi p}{\gcd(n_1, n_2, n_3, n_4)} \int_{S^3} \sum_{i,j,k,l}\delta(S^1_i\cap S^2_j\cap S^3_k \cap \partial^{-1}S^4_l)\bigg)|\mathcal{C}\rangle.
\end{eqnarray}
\end{widetext}

To compute the entanglement entropy, we first choose an entanglement cut, which bipartites the space into two parts. The ground state wavefunction factorizes accordingly,
\begin{eqnarray}\label{WF02}
|\psi\rangle= \mathfrak{C}\sum_{\mathcal{C}_A\in \mathcal{H}_A, \mathcal{C}_B\in \mathcal{H}_B}p(\mathcal{C}_A,\mathcal{C}_B)|\mathcal{C}_A\rangle\otimes |\mathcal{C}_B\rangle
\end{eqnarray}
where $p(\mathcal{C}_A,\mathcal{C}_B)=\exp( i \frac{2\pi p}{\gcd(n_1, n_2, n_3, n_4)} \int_{S^3} \sum_{i,j,k,l}
\delta(S^1_i\cap S^2_j\cap S^3_k \cap \partial^{-1}S^4_l))$ is a $U(1)$ factor. We would like to see whether the twisting
 parameter $p$ can affect the entanglement entropy. One crucial observation from Ref.\onlinecite{Zheng2017Structure} is
that unless the coefficient $p(\mathcal{C}_A,\mathcal{C}_B)$ has a term which can factorize into the product of the
contribution from the region $A$ and the contribution from the region $B$,
\textit{i.e.}, $\exp(i q(\{S_i\}|_{A})q'(\{S_j\}|_{B}))$ (where $q(\{S\}|_A)$ is an arbitrary quantity depending only on the
 configuration $\{S\}$ inside region $A$, and similarly for $q'(\{S\}|_B)$ ), the $U(1)$ factor can always be absorbed by
redefining the basis $|\mathcal{C}_A\rangle$ and $ |\mathcal{C}_B\rangle$.  However, from the form of $p(\mathcal{C}_A,
\mathcal{C}_B)$ (see Eq.~~\eqref{WFof IVDW}), to have nontrivial intersection, there must be at least one $S_i$ in
 $\delta(S^1_i\cap S^2_j\cap S^3_k \cap \partial^{-1}S^4_l)$ such that it intersects with the entanglement surface.
Hence the phase factor in Eq.~~\eqref{WFof IVDW} can be absorbed by redefinition of the states $|\mathcal{C}\rangle$
and the singular values of the resulting wavefunction is the same as the singular values of the wavefunction of pure
 $\otimes_{I=1}^4 \mathbb{Z}_{n_I}$ gauge theory without twisting terms.

Since it is well known that for an untwisted gauge theory, the entanglement entropy across $S^2$ and $T^2$ are the same. As a corollary of the above argument, we have $S_{\mathrm{topo}}[S^2]=S_{\mathrm{topo}}[T^2]$ for a type IV DW theory.

\section{Dimension reduction of cocycles from a slant product}
\label{Sec: 2cocycle}

In the main text, the 2-cocycles $\beta_{a,b}(c,d)$ appear in two different places, one
in the definition of irreducible projective representation in Eq.~\eqref{PRtype4}, and the other
in evolving from one-cube basis to two-cube basis in Eq.~\eqref{1to2}.
In this appendix, we give a brief introduction on slant production to obtain 2-cocycles from 4-cocycles.
For more details on group cohomology and cocycles, one may refer to, \textit{e.g.}, Refs.\onlinecite{JW1404,Wan1409,Jiang2014Generalized}.

For the dimension reduction of cochains, one can use slant product to map a $n$-cochain $c$ to $(n-1)$-cochain $i_gc$ as follows
\be\label{SlantProduct}
\begin{split}
&i_gc(g_1,g_2,\cdots,g_{n-1})=c(g,g_1,g_2,\cdots,g_{n-1})^{(-1)^{n-1}}\cdot
\prod_{j=1}^{n-1}\\
&c\big(g_1,\cdots,g_j,(g_1\cdots g_j)^{-1}\cdot g\cdot
(g_1\cdots g_j),\cdots,g_{n-1}\big)^{(-1)^{n-1+j}}
\end{split}
\ee
In this work, since we are working in (3+1)D, the theory will be defined based on the 4-cocycle
$\omega\in\mathcal{H}^4[G;U(1)]$. We use the `canonical' 4-cocycle so that $\omega(g_1,g_2,g_3,g_4)=1$
as long as any of $g_1,g_2,g_3$ and $g_4$ is $\mathbb 1$, the identity element of group $G$.
Then by doing slant product, we can obtain 3-cocycles from 4-cocycles:
\be\label{3cocycle}
\alpha_a(b,c,d)=i_a\omega(b,c,d).
\ee
From Eq.~\eqref{SlantProduct}, one can find that if $a=\mathbb 1$, then the
induced three-cocycle $\alpha_a(b,c,d)=1$, which is trivial.
Similarly, one can obtain the 2-cocycle from 3-cocycle:
\be\label{2cocycle}
\beta_{a,b}(c,d)=i_b \alpha_a (c,d)
=\frac{
\alpha_a(b,c,d) \alpha_a(c,d,b)
}{\alpha_a(c,b,d)},
\ee
if any of $a,b$ is equal to $\mathbb 1$, then $\beta_{a,b}(c,d)$ will be trivial, \textit{i.e.}, $\beta_{a,b}(c,d)=1$.
It is noted that the irreducible projective representation is determined by (see Eq.~\eqref{PRtype4})
\be
\tilde{\rho}^{a,b}_{\mu}(c)\tilde{\rho}^{a,b}_{\mu}(d)
=
\beta_{a,b}(c,d)
\tilde{\rho}^{a,b}_{\mu}(c\cdot d).
\ee
Then $\beta_{a,b}(c,d)=1$ indicates that the projective representation $\tilde{\rho}^{a,b}_{\mu}$
will reduce to the linear representation that satisfies $\rho^{a,b}_{\mu}(c)\rho^{a,b}_{\mu}(d)
=\rho^{a,b}_{\mu}(c\cdot d)$.
For the point-like excitations ($a=b=\mathbb 1$) and single-loop excitations (one of $a,b$ is equal to $\mathbb 1$),
one has $\beta_{a,b}(c,d)=1$. Therefore, these two types of excitations only carry linear representations,
and are insensitive to the 4-cocycles $\omega$.
On the other hand, for a Hopf-link loop excitation, one has $a\neq \mathbb 1$ and $b\neq \mathbb 1$, which may lead
to a nontrivial $\beta_{a,b}(c,d)$ and therefore a projective representation.

For an Abelian gauge group $G$ as studied in the main text, based on Eq.~\eqref{3cocycle}, one can obtain
\be\label{Abelian3cocycle}
\alpha_a(b,c,d)=\frac{
\omega(b,a,c,d)\omega(b,c,d,a)
}{
\omega(a,b,c,d)\omega(b,c,a,d)
}.
\ee
Then, based on Eqs.\eqref{2cocycle} and \eqref{Abelian3cocycle}, one can express the 2-cocycle
$\beta_{a,b}(c,d)$ in terms of 4-cocycles $\omega$.

Interestingly, there is a geometrical meaning of this 2-cocycle $\beta_{a,b}(c,d)$.
For example, $\beta_{a,b}(c_1,c_2)$ can be viewed as the phase associated with
the manifold in Fig.~\ref{Triangulation}\,(b), after an appropriate coloring of the manifold.
We refer the readers to Ref.\onlinecite{Jiang2014Generalized} for more details on this geometrical picture.

\bibliography{NonAbelianEntanglement}

\begin{thebibliography}{113}%
\makeatletter
\providecommand \@ifxundefined [1]{%
 \@ifx{#1\undefined}
}%
\providecommand \@ifnum [1]{%
 \ifnum #1\expandafter \@firstoftwo
 \else \expandafter \@secondoftwo
 \fi
}%
\providecommand \@ifx [1]{%
 \ifx #1\expandafter \@firstoftwo
 \else \expandafter \@secondoftwo
 \fi
}%
\providecommand \natexlab [1]{#1}%
\providecommand \enquote  [1]{``#1''}%
\providecommand \bibnamefont  [1]{#1}%
\providecommand \bibfnamefont [1]{#1}%
\providecommand \citenamefont [1]{#1}%
\providecommand \href@noop [0]{\@secondoftwo}%
\providecommand \href [0]{\begingroup \@sanitize@url \@href}%
\providecommand \@href[1]{\@@startlink{#1}\@@href}%
\providecommand \@@href[1]{\endgroup#1\@@endlink}%
\providecommand \@sanitize@url [0]{\catcode `\\12\catcode `\$12\catcode
  `\&12\catcode `\#12\catcode `\^12\catcode `\_12\catcode `\%12\relax}%
\providecommand \@@startlink[1]{}%
\providecommand \@@endlink[0]{}%
\providecommand \url  [0]{\begingroup\@sanitize@url \@url }%
\providecommand \@url [1]{\endgroup\@href {#1}{\urlprefix }}%
\providecommand \urlprefix  [0]{URL }%
\providecommand \Eprint [0]{\href }%
\providecommand \doibase [0]{http://dx.doi.org/}%
\providecommand \selectlanguage [0]{\@gobble}%
\providecommand \bibinfo  [0]{\@secondoftwo}%
\providecommand \bibfield  [0]{\@secondoftwo}%
\providecommand \translation [1]{[#1]}%
\providecommand \BibitemOpen [0]{}%
\providecommand \bibitemStop [0]{}%
\providecommand \bibitemNoStop [0]{.\EOS\space}%
\providecommand \EOS [0]{\spacefactor3000\relax}%
\providecommand \BibitemShut  [1]{\csname bibitem#1\endcsname}%
\let\auto@bib@innerbib\@empty
\bibitem [{\citenamefont {Wen}(2004)}]{Wen04book}%
  \BibitemOpen
  \bibfield  {author} {\bibinfo {author} {\bibfnamefont {X.-G.}\ \bibnamefont
  {Wen}},\ }\href@noop {} {\enquote {\bibinfo {title} {Quantum field theory of
  many-body systems},}\ } (\bibinfo {year} {2004})\BibitemShut {NoStop}%
\bibitem [{\citenamefont {Arovas}\ \emph {et~al.}(1984)\citenamefont {Arovas},
  \citenamefont {Schrieffer},\ and\ \citenamefont {Wilczek}}]{Wilczek1984}%
  \BibitemOpen
  \bibfield  {author} {\bibinfo {author} {\bibfnamefont {D.}~\bibnamefont
  {Arovas}}, \bibinfo {author} {\bibfnamefont {J.~R.}\ \bibnamefont
  {Schrieffer}}, \ and\ \bibinfo {author} {\bibfnamefont {F.}~\bibnamefont
  {Wilczek}},\ }\href {\doibase 10.1103/PhysRevLett.53.722} {\bibfield
  {journal} {\bibinfo  {journal} {Phys. Rev. Lett.}\ }\textbf {\bibinfo
  {volume} {53}},\ \bibinfo {pages} {722} (\bibinfo {year} {1984})}\BibitemShut
  {NoStop}%
\bibitem [{\citenamefont {Wen}\ and\ \citenamefont {Niu}(1990)}]{Wen1990}%
  \BibitemOpen
  \bibfield  {author} {\bibinfo {author} {\bibfnamefont {X.~G.}\ \bibnamefont
  {Wen}}\ and\ \bibinfo {author} {\bibfnamefont {Q.}~\bibnamefont {Niu}},\
  }\href {\doibase 10.1103/PhysRevB.41.9377} {\bibfield  {journal} {\bibinfo
  {journal} {Phys. Rev. B}\ }\textbf {\bibinfo {volume} {41}},\ \bibinfo
  {pages} {9377} (\bibinfo {year} {1990})}\BibitemShut {NoStop}%
\bibitem [{\citenamefont {Levin}\ and\ \citenamefont {Wen}(2006)}]{levin2006}%
  \BibitemOpen
  \bibfield  {author} {\bibinfo {author} {\bibfnamefont {M.}~\bibnamefont
  {Levin}}\ and\ \bibinfo {author} {\bibfnamefont {X.-G.}\ \bibnamefont
  {Wen}},\ }\href {\doibase 10.1103/PhysRevLett.96.110405} {\bibfield
  {journal} {\bibinfo  {journal} {Phys. Rev. Lett.}\ }\textbf {\bibinfo
  {volume} {96}},\ \bibinfo {pages} {110405} (\bibinfo {year}
  {2006})}\BibitemShut {NoStop}%
\bibitem [{\citenamefont {Kitaev}\ and\ \citenamefont
  {Preskill}(2006)}]{kitaev2006}%
  \BibitemOpen
  \bibfield  {author} {\bibinfo {author} {\bibfnamefont {A.}~\bibnamefont
  {Kitaev}}\ and\ \bibinfo {author} {\bibfnamefont {J.}~\bibnamefont
  {Preskill}},\ }\href {\doibase 10.1103/PhysRevLett.96.110404} {\bibfield
  {journal} {\bibinfo  {journal} {Phys. Rev. Lett.}\ }\textbf {\bibinfo
  {volume} {96}},\ \bibinfo {pages} {110404} (\bibinfo {year}
  {2006})}\BibitemShut {NoStop}%
\bibitem [{\citenamefont {Li}\ and\ \citenamefont
  {Haldane}(2008)}]{Li2008Entanglement}%
  \BibitemOpen
  \bibfield  {author} {\bibinfo {author} {\bibfnamefont {H.}~\bibnamefont
  {Li}}\ and\ \bibinfo {author} {\bibfnamefont {F.~D.~M.}\ \bibnamefont
  {Haldane}},\ }\href {\doibase 10.1103/PhysRevLett.101.010504} {\bibfield
  {journal} {\bibinfo  {journal} {Phys. Rev. Lett.}\ }\textbf {\bibinfo
  {volume} {101}},\ \bibinfo {pages} {010504} (\bibinfo {year}
  {2008})}\BibitemShut {NoStop}%
\bibitem [{\citenamefont {Dong}\ \emph {et~al.}(2008)\citenamefont {Dong},
  \citenamefont {Fradkin}, \citenamefont {Leigh},\ and\ \citenamefont
  {Nowling}}]{Dong0802}%
  \BibitemOpen
  \bibfield  {author} {\bibinfo {author} {\bibfnamefont {S.}~\bibnamefont
  {Dong}}, \bibinfo {author} {\bibfnamefont {E.}~\bibnamefont {Fradkin}},
  \bibinfo {author} {\bibfnamefont {R.~G.}\ \bibnamefont {Leigh}}, \ and\
  \bibinfo {author} {\bibfnamefont {S.}~\bibnamefont {Nowling}},\ }\href@noop
  {} {\bibfield  {journal} {\bibinfo  {journal} {Journal of High Energy
  Physics}\ }\textbf {\bibinfo {volume} {2008}},\ \bibinfo {pages} {016}
  (\bibinfo {year} {2008})}\BibitemShut {NoStop}%
\bibitem [{\citenamefont {Zhang}\ \emph {et~al.}(2012)\citenamefont {Zhang},
  \citenamefont {Grover}, \citenamefont {Turner}, \citenamefont {Oshikawa},\
  and\ \citenamefont {Vishwanath}}]{zhang2012}%
  \BibitemOpen
  \bibfield  {author} {\bibinfo {author} {\bibfnamefont {Y.}~\bibnamefont
  {Zhang}}, \bibinfo {author} {\bibfnamefont {T.}~\bibnamefont {Grover}},
  \bibinfo {author} {\bibfnamefont {A.}~\bibnamefont {Turner}}, \bibinfo
  {author} {\bibfnamefont {M.}~\bibnamefont {Oshikawa}}, \ and\ \bibinfo
  {author} {\bibfnamefont {A.}~\bibnamefont {Vishwanath}},\ }\href {\doibase
  10.1103/PhysRevB.85.235151} {\bibfield  {journal} {\bibinfo  {journal} {Phys.
  Rev. B}\ }\textbf {\bibinfo {volume} {85}},\ \bibinfo {pages} {235151}
  (\bibinfo {year} {2012})}\BibitemShut {NoStop}%
\bibitem [{\citenamefont {Regnault}\ \emph {et~al.}(2009)\citenamefont
  {Regnault}, \citenamefont {Bernevig},\ and\ \citenamefont
  {Haldane}}]{Regnault2009Topological}%
  \BibitemOpen
  \bibfield  {author} {\bibinfo {author} {\bibfnamefont {N.}~\bibnamefont
  {Regnault}}, \bibinfo {author} {\bibfnamefont {B.~A.}\ \bibnamefont
  {Bernevig}}, \ and\ \bibinfo {author} {\bibfnamefont {F.~D.~M.}\ \bibnamefont
  {Haldane}},\ }\href {\doibase 10.1103/PhysRevLett.103.016801} {\bibfield
  {journal} {\bibinfo  {journal} {Phys. Rev. Lett.}\ }\textbf {\bibinfo
  {volume} {103}},\ \bibinfo {pages} {016801} (\bibinfo {year}
  {2009})}\BibitemShut {NoStop}%
\bibitem [{\citenamefont {Thomale}\ \emph
  {et~al.}(2010{\natexlab{a}})\citenamefont {Thomale}, \citenamefont {Arovas},\
  and\ \citenamefont {Bernevig}}]{Thomale2010Nonlocal}%
  \BibitemOpen
  \bibfield  {author} {\bibinfo {author} {\bibfnamefont {R.}~\bibnamefont
  {Thomale}}, \bibinfo {author} {\bibfnamefont {D.~P.}\ \bibnamefont {Arovas}},
  \ and\ \bibinfo {author} {\bibfnamefont {B.~A.}\ \bibnamefont {Bernevig}},\
  }\href {\doibase 10.1103/PhysRevLett.105.116805} {\bibfield  {journal}
  {\bibinfo  {journal} {Phys. Rev. Lett.}\ }\textbf {\bibinfo {volume} {105}},\
  \bibinfo {pages} {116805} (\bibinfo {year} {2010}{\natexlab{a}})}\BibitemShut
  {NoStop}%
\bibitem [{\citenamefont {Thomale}\ \emph
  {et~al.}(2010{\natexlab{b}})\citenamefont {Thomale}, \citenamefont
  {Sterdyniak}, \citenamefont {Regnault},\ and\ \citenamefont
  {Bernevig}}]{Thomale2010Entanglement}%
  \BibitemOpen
  \bibfield  {author} {\bibinfo {author} {\bibfnamefont {R.}~\bibnamefont
  {Thomale}}, \bibinfo {author} {\bibfnamefont {A.}~\bibnamefont {Sterdyniak}},
  \bibinfo {author} {\bibfnamefont {N.}~\bibnamefont {Regnault}}, \ and\
  \bibinfo {author} {\bibfnamefont {B.~A.}\ \bibnamefont {Bernevig}},\ }\href
  {\doibase 10.1103/PhysRevLett.104.180502} {\bibfield  {journal} {\bibinfo
  {journal} {Phys. Rev. Lett.}\ }\textbf {\bibinfo {volume} {104}},\ \bibinfo
  {pages} {180502} (\bibinfo {year} {2010}{\natexlab{b}})}\BibitemShut
  {NoStop}%
\bibitem [{\citenamefont {Prodan}\ \emph {et~al.}(2010)\citenamefont {Prodan},
  \citenamefont {Hughes},\ and\ \citenamefont
  {Bernevig}}]{Prodan2010Entanglement}%
  \BibitemOpen
  \bibfield  {author} {\bibinfo {author} {\bibfnamefont {E.}~\bibnamefont
  {Prodan}}, \bibinfo {author} {\bibfnamefont {T.~L.}\ \bibnamefont {Hughes}},
  \ and\ \bibinfo {author} {\bibfnamefont {B.~A.}\ \bibnamefont {Bernevig}},\
  }\href {\doibase 10.1103/PhysRevLett.105.115501} {\bibfield  {journal}
  {\bibinfo  {journal} {Phys. Rev. Lett.}\ }\textbf {\bibinfo {volume} {105}},\
  \bibinfo {pages} {115501} (\bibinfo {year} {2010})}\BibitemShut {NoStop}%
\bibitem [{\citenamefont {Sterdyniak}\ \emph
  {et~al.}(2011{\natexlab{a}})\citenamefont {Sterdyniak}, \citenamefont
  {Regnault},\ and\ \citenamefont {Bernevig}}]{Sterdyniak2011Extracting}%
  \BibitemOpen
  \bibfield  {author} {\bibinfo {author} {\bibfnamefont {A.}~\bibnamefont
  {Sterdyniak}}, \bibinfo {author} {\bibfnamefont {N.}~\bibnamefont
  {Regnault}}, \ and\ \bibinfo {author} {\bibfnamefont {B.~A.}\ \bibnamefont
  {Bernevig}},\ }\href {\doibase 10.1103/PhysRevLett.106.100405} {\bibfield
  {journal} {\bibinfo  {journal} {Phys. Rev. Lett.}\ }\textbf {\bibinfo
  {volume} {106}},\ \bibinfo {pages} {100405} (\bibinfo {year}
  {2011}{\natexlab{a}})}\BibitemShut {NoStop}%
\bibitem [{\citenamefont {Papi\ifmmode~\acute{c}\else \'{c}\fi{}}\ \emph
  {et~al.}(2011)\citenamefont {Papi\ifmmode~\acute{c}\else \'{c}\fi{}},
  \citenamefont {Bernevig},\ and\ \citenamefont
  {Regnault}}]{Papic2011Topological}%
  \BibitemOpen
  \bibfield  {author} {\bibinfo {author} {\bibfnamefont {Z.}~\bibnamefont
  {Papi\ifmmode~\acute{c}\else \'{c}\fi{}}}, \bibinfo {author} {\bibfnamefont
  {B.~A.}\ \bibnamefont {Bernevig}}, \ and\ \bibinfo {author} {\bibfnamefont
  {N.}~\bibnamefont {Regnault}},\ }\href {\doibase
  10.1103/PhysRevLett.106.056801} {\bibfield  {journal} {\bibinfo  {journal}
  {Phys. Rev. Lett.}\ }\textbf {\bibinfo {volume} {106}},\ \bibinfo {pages}
  {056801} (\bibinfo {year} {2011})}\BibitemShut {NoStop}%
\bibitem [{\citenamefont {Hermanns}\ \emph {et~al.}(2011)\citenamefont
  {Hermanns}, \citenamefont {Chandran}, \citenamefont {Regnault},\ and\
  \citenamefont {Bernevig}}]{Hermanns2011Haldane}%
  \BibitemOpen
  \bibfield  {author} {\bibinfo {author} {\bibfnamefont {M.}~\bibnamefont
  {Hermanns}}, \bibinfo {author} {\bibfnamefont {A.}~\bibnamefont {Chandran}},
  \bibinfo {author} {\bibfnamefont {N.}~\bibnamefont {Regnault}}, \ and\
  \bibinfo {author} {\bibfnamefont {B.~A.}\ \bibnamefont {Bernevig}},\ }\href
  {\doibase 10.1103/PhysRevB.84.121309} {\bibfield  {journal} {\bibinfo
  {journal} {Phys. Rev. B}\ }\textbf {\bibinfo {volume} {84}},\ \bibinfo
  {pages} {121309} (\bibinfo {year} {2011})}\BibitemShut {NoStop}%
\bibitem [{\citenamefont {Chandran}\ \emph {et~al.}(2011)\citenamefont
  {Chandran}, \citenamefont {Hermanns}, \citenamefont {Regnault},\ and\
  \citenamefont {Bernevig}}]{Chandran2011Bulk}%
  \BibitemOpen
  \bibfield  {author} {\bibinfo {author} {\bibfnamefont {A.}~\bibnamefont
  {Chandran}}, \bibinfo {author} {\bibfnamefont {M.}~\bibnamefont {Hermanns}},
  \bibinfo {author} {\bibfnamefont {N.}~\bibnamefont {Regnault}}, \ and\
  \bibinfo {author} {\bibfnamefont {B.~A.}\ \bibnamefont {Bernevig}},\ }\href
  {\doibase 10.1103/PhysRevB.84.205136} {\bibfield  {journal} {\bibinfo
  {journal} {Phys. Rev. B}\ }\textbf {\bibinfo {volume} {84}},\ \bibinfo
  {pages} {205136} (\bibinfo {year} {2011})}\BibitemShut {NoStop}%
\bibitem [{\citenamefont {Regnault}\ and\ \citenamefont
  {Bernevig}(2011)}]{Regnault2011Fractional}%
  \BibitemOpen
  \bibfield  {author} {\bibinfo {author} {\bibfnamefont {N.}~\bibnamefont
  {Regnault}}\ and\ \bibinfo {author} {\bibfnamefont {B.~A.}\ \bibnamefont
  {Bernevig}},\ }\href {\doibase 10.1103/PhysRevX.1.021014} {\bibfield
  {journal} {\bibinfo  {journal} {Phys. Rev. X}\ }\textbf {\bibinfo {volume}
  {1}},\ \bibinfo {pages} {021014} (\bibinfo {year} {2011})}\BibitemShut
  {NoStop}%
\bibitem [{\citenamefont {Sterdyniak}\ \emph
  {et~al.}(2011{\natexlab{b}})\citenamefont {Sterdyniak}, \citenamefont
  {Bernevig}, \citenamefont {Regnault},\ and\ \citenamefont
  {Haldane}}]{Sterdyniak2011Hierarchical}%
  \BibitemOpen
  \bibfield  {author} {\bibinfo {author} {\bibfnamefont {A.}~\bibnamefont
  {Sterdyniak}}, \bibinfo {author} {\bibfnamefont {B.~A.}\ \bibnamefont
  {Bernevig}}, \bibinfo {author} {\bibfnamefont {N.}~\bibnamefont {Regnault}},
  \ and\ \bibinfo {author} {\bibfnamefont {F.~D.~M.}\ \bibnamefont {Haldane}},\
  }\href {http://stacks.iop.org/1367-2630/13/i=10/a=105001} {\bibfield
  {journal} {\bibinfo  {journal} {New Journal of Physics}\ }\textbf {\bibinfo
  {volume} {13}},\ \bibinfo {pages} {105001} (\bibinfo {year}
  {2011}{\natexlab{b}})}\BibitemShut {NoStop}%
\bibitem [{\citenamefont {Alexandradinata}\ \emph {et~al.}(2011)\citenamefont
  {Alexandradinata}, \citenamefont {Hughes},\ and\ \citenamefont
  {Bernevig}}]{Alexandradinata2011Trace}%
  \BibitemOpen
  \bibfield  {author} {\bibinfo {author} {\bibfnamefont {A.}~\bibnamefont
  {Alexandradinata}}, \bibinfo {author} {\bibfnamefont {T.~L.}\ \bibnamefont
  {Hughes}}, \ and\ \bibinfo {author} {\bibfnamefont {B.~A.}\ \bibnamefont
  {Bernevig}},\ }\href {\doibase 10.1103/PhysRevB.84.195103} {\bibfield
  {journal} {\bibinfo  {journal} {Phys. Rev. B}\ }\textbf {\bibinfo {volume}
  {84}},\ \bibinfo {pages} {195103} (\bibinfo {year} {2011})}\BibitemShut
  {NoStop}%
\bibitem [{\citenamefont {Sterdyniak}\ \emph {et~al.}(2012)\citenamefont
  {Sterdyniak}, \citenamefont {Chandran}, \citenamefont {Regnault},
  \citenamefont {Bernevig},\ and\ \citenamefont
  {Bonderson}}]{Sterdyniak2012Real}%
  \BibitemOpen
  \bibfield  {author} {\bibinfo {author} {\bibfnamefont {A.}~\bibnamefont
  {Sterdyniak}}, \bibinfo {author} {\bibfnamefont {A.}~\bibnamefont
  {Chandran}}, \bibinfo {author} {\bibfnamefont {N.}~\bibnamefont {Regnault}},
  \bibinfo {author} {\bibfnamefont {B.~A.}\ \bibnamefont {Bernevig}}, \ and\
  \bibinfo {author} {\bibfnamefont {P.}~\bibnamefont {Bonderson}},\ }\href
  {\doibase 10.1103/PhysRevB.85.125308} {\bibfield  {journal} {\bibinfo
  {journal} {Phys. Rev. B}\ }\textbf {\bibinfo {volume} {85}},\ \bibinfo
  {pages} {125308} (\bibinfo {year} {2012})}\BibitemShut {NoStop}%
\bibitem [{\citenamefont {Gilbert}\ \emph {et~al.}(2012)\citenamefont
  {Gilbert}, \citenamefont {Bernevig},\ and\ \citenamefont
  {Hughes}}]{Gilbert2012Signature}%
  \BibitemOpen
  \bibfield  {author} {\bibinfo {author} {\bibfnamefont {M.~J.}\ \bibnamefont
  {Gilbert}}, \bibinfo {author} {\bibfnamefont {B.~A.}\ \bibnamefont
  {Bernevig}}, \ and\ \bibinfo {author} {\bibfnamefont {T.~L.}\ \bibnamefont
  {Hughes}},\ }\href {\doibase 10.1103/PhysRevB.86.041401} {\bibfield
  {journal} {\bibinfo  {journal} {Phys. Rev. B}\ }\textbf {\bibinfo {volume}
  {86}},\ \bibinfo {pages} {041401} (\bibinfo {year} {2012})}\BibitemShut
  {NoStop}%
\bibitem [{\citenamefont {Sterdyniak}\ \emph {et~al.}(2013)\citenamefont
  {Sterdyniak}, \citenamefont {Repellin}, \citenamefont {Bernevig},\ and\
  \citenamefont {Regnault}}]{Sterdyniak2013Series}%
  \BibitemOpen
  \bibfield  {author} {\bibinfo {author} {\bibfnamefont {A.}~\bibnamefont
  {Sterdyniak}}, \bibinfo {author} {\bibfnamefont {C.}~\bibnamefont
  {Repellin}}, \bibinfo {author} {\bibfnamefont {B.~A.}\ \bibnamefont
  {Bernevig}}, \ and\ \bibinfo {author} {\bibfnamefont {N.}~\bibnamefont
  {Regnault}},\ }\href {\doibase 10.1103/PhysRevB.87.205137} {\bibfield
  {journal} {\bibinfo  {journal} {Phys. Rev. B}\ }\textbf {\bibinfo {volume}
  {87}},\ \bibinfo {pages} {205137} (\bibinfo {year} {2013})}\BibitemShut
  {NoStop}%
\bibitem [{\citenamefont {Estienne}\ \emph {et~al.}(2015)\citenamefont
  {Estienne}, \citenamefont {Regnault},\ and\ \citenamefont
  {Bernevig}}]{Estienne2015Correlation}%
  \BibitemOpen
  \bibfield  {author} {\bibinfo {author} {\bibfnamefont {B.}~\bibnamefont
  {Estienne}}, \bibinfo {author} {\bibfnamefont {N.}~\bibnamefont {Regnault}},
  \ and\ \bibinfo {author} {\bibfnamefont {B.~A.}\ \bibnamefont {Bernevig}},\
  }\href {\doibase 10.1103/PhysRevLett.114.186801} {\bibfield  {journal}
  {\bibinfo  {journal} {Phys. Rev. Lett.}\ }\textbf {\bibinfo {volume} {114}},\
  \bibinfo {pages} {186801} (\bibinfo {year} {2015})}\BibitemShut {NoStop}%
\bibitem [{\citenamefont {Wen}\ \emph {et~al.}(2016)\citenamefont {Wen},
  \citenamefont {Matsuura},\ and\ \citenamefont {Ryu}}]{Wen2016Edge}%
  \BibitemOpen
  \bibfield  {author} {\bibinfo {author} {\bibfnamefont {X.}~\bibnamefont
  {Wen}}, \bibinfo {author} {\bibfnamefont {S.}~\bibnamefont {Matsuura}}, \
  and\ \bibinfo {author} {\bibfnamefont {S.}~\bibnamefont {Ryu}},\ }\href
  {\doibase 10.1103/PhysRevB.93.245140} {\bibfield  {journal} {\bibinfo
  {journal} {Phys. Rev. B}\ }\textbf {\bibinfo {volume} {93}},\ \bibinfo
  {pages} {245140} (\bibinfo {year} {2016})}\BibitemShut {NoStop}%
\bibitem [{\citenamefont {He}\ \emph {et~al.}(2014)\citenamefont {He},
  \citenamefont {Moradi},\ and\ \citenamefont {Wen}}]{He2014Modular}%
  \BibitemOpen
  \bibfield  {author} {\bibinfo {author} {\bibfnamefont {H.}~\bibnamefont
  {He}}, \bibinfo {author} {\bibfnamefont {H.}~\bibnamefont {Moradi}}, \ and\
  \bibinfo {author} {\bibfnamefont {X.-G.}\ \bibnamefont {Wen}},\ }\href
  {\doibase 10.1103/PhysRevB.90.205114} {\bibfield  {journal} {\bibinfo
  {journal} {Phys. Rev. B}\ }\textbf {\bibinfo {volume} {90}},\ \bibinfo
  {pages} {205114} (\bibinfo {year} {2014})}\BibitemShut {NoStop}%
\bibitem [{\citenamefont {Moradi}\ and\ \citenamefont
  {Wen}(2015)}]{Moradi1404}%
  \BibitemOpen
  \bibfield  {author} {\bibinfo {author} {\bibfnamefont {H.}~\bibnamefont
  {Moradi}}\ and\ \bibinfo {author} {\bibfnamefont {X.-G.}\ \bibnamefont
  {Wen}},\ }\href {\doibase 10.1103/PhysRevB.91.075114} {\bibfield  {journal}
  {\bibinfo  {journal} {Phys. Rev. B}\ }\textbf {\bibinfo {volume} {91}},\
  \bibinfo {pages} {075114} (\bibinfo {year} {2015})}\BibitemShut {NoStop}%
\bibitem [{\citenamefont {Mei}\ and\ \citenamefont
  {Wen}(2015)}]{Mei2015Modular}%
  \BibitemOpen
  \bibfield  {author} {\bibinfo {author} {\bibfnamefont {J.-W.}\ \bibnamefont
  {Mei}}\ and\ \bibinfo {author} {\bibfnamefont {X.-G.}\ \bibnamefont {Wen}},\
  }\href {\doibase 10.1103/PhysRevB.91.125123} {\bibfield  {journal} {\bibinfo
  {journal} {Phys. Rev. B}\ }\textbf {\bibinfo {volume} {91}},\ \bibinfo
  {pages} {125123} (\bibinfo {year} {2015})}\BibitemShut {NoStop}%
\bibitem [{\citenamefont {Zhang}\ \emph {et~al.}(2015)\citenamefont {Zhang},
  \citenamefont {Grover},\ and\ \citenamefont {Vishwanath}}]{Zhang2015General}%
  \BibitemOpen
  \bibfield  {author} {\bibinfo {author} {\bibfnamefont {Y.}~\bibnamefont
  {Zhang}}, \bibinfo {author} {\bibfnamefont {T.}~\bibnamefont {Grover}}, \
  and\ \bibinfo {author} {\bibfnamefont {A.}~\bibnamefont {Vishwanath}},\
  }\href {\doibase 10.1103/PhysRevB.91.035127} {\bibfield  {journal} {\bibinfo
  {journal} {Phys. Rev. B}\ }\textbf {\bibinfo {volume} {91}},\ \bibinfo
  {pages} {035127} (\bibinfo {year} {2015})}\BibitemShut {NoStop}%
\bibitem [{\citenamefont {You}\ and\ \citenamefont
  {Cheng}(2015)}]{you2015measuring}%
  \BibitemOpen
  \bibfield  {author} {\bibinfo {author} {\bibfnamefont {Y.-Z.}\ \bibnamefont
  {You}}\ and\ \bibinfo {author} {\bibfnamefont {M.}~\bibnamefont {Cheng}},\
  }\href@noop {} {\bibfield  {journal} {\bibinfo  {journal} {arXiv preprint
  arXiv:1502.03192}\ } (\bibinfo {year} {2015})}\BibitemShut {NoStop}%
\bibitem [{\citenamefont {Huang}\ and\ \citenamefont
  {Wei}(2015)}]{Huang2015Transition}%
  \BibitemOpen
  \bibfield  {author} {\bibinfo {author} {\bibfnamefont {C.-Y.}\ \bibnamefont
  {Huang}}\ and\ \bibinfo {author} {\bibfnamefont {T.-C.}\ \bibnamefont
  {Wei}},\ }\href {\doibase 10.1103/PhysRevB.92.085405} {\bibfield  {journal}
  {\bibinfo  {journal} {Phys. Rev. B}\ }\textbf {\bibinfo {volume} {92}},\
  \bibinfo {pages} {085405} (\bibinfo {year} {2015})}\BibitemShut {NoStop}%
\bibitem [{\citenamefont {Huang}\ and\ \citenamefont
  {Wei}(2016)}]{Huang2016Detecting}%
  \BibitemOpen
  \bibfield  {author} {\bibinfo {author} {\bibfnamefont {C.-Y.}\ \bibnamefont
  {Huang}}\ and\ \bibinfo {author} {\bibfnamefont {T.-C.}\ \bibnamefont
  {Wei}},\ }\href {\doibase 10.1103/PhysRevB.93.155163} {\bibfield  {journal}
  {\bibinfo  {journal} {Phys. Rev. B}\ }\textbf {\bibinfo {volume} {93}},\
  \bibinfo {pages} {155163} (\bibinfo {year} {2016})}\BibitemShut {NoStop}%
\bibitem [{\citenamefont {Li}\ \emph {et~al.}(2017)\citenamefont {Li},
  \citenamefont {Wan}, \citenamefont {Hung}, \citenamefont {Lan}, \citenamefont
  {Long}, \citenamefont {Lu}, \citenamefont {Zeng},\ and\ \citenamefont
  {Laflamme}}]{Li2017Experimental}%
  \BibitemOpen
  \bibfield  {author} {\bibinfo {author} {\bibfnamefont {K.}~\bibnamefont
  {Li}}, \bibinfo {author} {\bibfnamefont {Y.}~\bibnamefont {Wan}}, \bibinfo
  {author} {\bibfnamefont {L.-Y.}\ \bibnamefont {Hung}}, \bibinfo {author}
  {\bibfnamefont {T.}~\bibnamefont {Lan}}, \bibinfo {author} {\bibfnamefont
  {G.}~\bibnamefont {Long}}, \bibinfo {author} {\bibfnamefont {D.}~\bibnamefont
  {Lu}}, \bibinfo {author} {\bibfnamefont {B.}~\bibnamefont {Zeng}}, \ and\
  \bibinfo {author} {\bibfnamefont {R.}~\bibnamefont {Laflamme}},\ }\href
  {\doibase 10.1103/PhysRevLett.118.080502} {\bibfield  {journal} {\bibinfo
  {journal} {Phys. Rev. Lett.}\ }\textbf {\bibinfo {volume} {118}},\ \bibinfo
  {pages} {080502} (\bibinfo {year} {2017})}\BibitemShut {NoStop}%
\bibitem [{\citenamefont {Mei}\ \emph {et~al.}(2017)\citenamefont {Mei},
  \citenamefont {Chen}, \citenamefont {He},\ and\ \citenamefont
  {Wen}}]{Mei2017Gapped}%
  \BibitemOpen
  \bibfield  {author} {\bibinfo {author} {\bibfnamefont {J.-W.}\ \bibnamefont
  {Mei}}, \bibinfo {author} {\bibfnamefont {J.-Y.}\ \bibnamefont {Chen}},
  \bibinfo {author} {\bibfnamefont {H.}~\bibnamefont {He}}, \ and\ \bibinfo
  {author} {\bibfnamefont {X.-G.}\ \bibnamefont {Wen}},\ }\href {\doibase
  10.1103/PhysRevB.95.235107} {\bibfield  {journal} {\bibinfo  {journal} {Phys.
  Rev. B}\ }\textbf {\bibinfo {volume} {95}},\ \bibinfo {pages} {235107}
  (\bibinfo {year} {2017})}\BibitemShut {NoStop}%
\bibitem [{\citenamefont {Walker}\ and\ \citenamefont
  {Wang}(2012)}]{Walker2012}%
  \BibitemOpen
  \bibfield  {author} {\bibinfo {author} {\bibfnamefont {K.}~\bibnamefont
  {Walker}}\ and\ \bibinfo {author} {\bibfnamefont {Z.}~\bibnamefont {Wang}},\
  }\href {\doibase 10.1007/s11467-011-0194-z} {\bibfield  {journal} {\bibinfo
  {journal} {Frontiers of Physics}\ }\textbf {\bibinfo {volume} {7}},\ \bibinfo
  {pages} {150} (\bibinfo {year} {2012})}\BibitemShut {NoStop}%
\bibitem [{\citenamefont {von Keyserlingk}\ \emph {et~al.}(2013)\citenamefont
  {von Keyserlingk}, \citenamefont {Burnell},\ and\ \citenamefont
  {Simon}}]{Curt2013}%
  \BibitemOpen
  \bibfield  {author} {\bibinfo {author} {\bibfnamefont {C.~W.}\ \bibnamefont
  {von Keyserlingk}}, \bibinfo {author} {\bibfnamefont {F.~J.}\ \bibnamefont
  {Burnell}}, \ and\ \bibinfo {author} {\bibfnamefont {S.~H.}\ \bibnamefont
  {Simon}},\ }\href {\doibase 10.1103/PhysRevB.87.045107} {\bibfield  {journal}
  {\bibinfo  {journal} {Phys. Rev. B}\ }\textbf {\bibinfo {volume} {87}},\
  \bibinfo {pages} {045107} (\bibinfo {year} {2013})}\BibitemShut {NoStop}%
\bibitem [{\citenamefont {Wang}\ and\ \citenamefont
  {Levin}(2014)}]{Wang2014Braiding}%
  \BibitemOpen
  \bibfield  {author} {\bibinfo {author} {\bibfnamefont {C.}~\bibnamefont
  {Wang}}\ and\ \bibinfo {author} {\bibfnamefont {M.}~\bibnamefont {Levin}},\
  }\href {\doibase 10.1103/PhysRevLett.113.080403} {\bibfield  {journal}
  {\bibinfo  {journal} {Phys. Rev. Lett.}\ }\textbf {\bibinfo {volume} {113}},\
  \bibinfo {pages} {080403} (\bibinfo {year} {2014})}\BibitemShut {NoStop}%
\bibitem [{\citenamefont {Jiang}\ \emph {et~al.}(2014)\citenamefont {Jiang},
  \citenamefont {Mesaros},\ and\ \citenamefont {Ran}}]{Jiang2014Generalized}%
  \BibitemOpen
  \bibfield  {author} {\bibinfo {author} {\bibfnamefont {S.}~\bibnamefont
  {Jiang}}, \bibinfo {author} {\bibfnamefont {A.}~\bibnamefont {Mesaros}}, \
  and\ \bibinfo {author} {\bibfnamefont {Y.}~\bibnamefont {Ran}},\ }\href
  {\doibase 10.1103/PhysRevX.4.031048} {\bibfield  {journal} {\bibinfo
  {journal} {Phys. Rev. X}\ }\textbf {\bibinfo {volume} {4}},\ \bibinfo {pages}
  {031048} (\bibinfo {year} {2014})}\BibitemShut {NoStop}%
\bibitem [{\citenamefont {Wang}\ and\ \citenamefont {Wen}(2015)}]{JW1404}%
  \BibitemOpen
  \bibfield  {author} {\bibinfo {author} {\bibfnamefont {J.~C.}\ \bibnamefont
  {Wang}}\ and\ \bibinfo {author} {\bibfnamefont {X.-G.}\ \bibnamefont {Wen}},\
  }\href {\doibase 10.1103/PhysRevB.91.035134} {\bibfield  {journal} {\bibinfo
  {journal} {Phys. Rev. B}\ }\textbf {\bibinfo {volume} {91}},\ \bibinfo
  {pages} {035134} (\bibinfo {year} {2015})}\BibitemShut {NoStop}%
\bibitem [{\citenamefont {Wan}\ \emph {et~al.}(2015)\citenamefont {Wan},
  \citenamefont {Wang},\ and\ \citenamefont {He}}]{Wan1409}%
  \BibitemOpen
  \bibfield  {author} {\bibinfo {author} {\bibfnamefont {Y.}~\bibnamefont
  {Wan}}, \bibinfo {author} {\bibfnamefont {J.~C.}\ \bibnamefont {Wang}}, \
  and\ \bibinfo {author} {\bibfnamefont {H.}~\bibnamefont {He}},\ }\href
  {\doibase 10.1103/PhysRevB.92.045101} {\bibfield  {journal} {\bibinfo
  {journal} {Phys. Rev. B}\ }\textbf {\bibinfo {volume} {92}},\ \bibinfo
  {pages} {045101} (\bibinfo {year} {2015})}\BibitemShut {NoStop}%
\bibitem [{\citenamefont {Bi}\ \emph {et~al.}(2014)\citenamefont {Bi},
  \citenamefont {You},\ and\ \citenamefont {Xu}}]{Bi2014Anyon}%
  \BibitemOpen
  \bibfield  {author} {\bibinfo {author} {\bibfnamefont {Z.}~\bibnamefont
  {Bi}}, \bibinfo {author} {\bibfnamefont {Y.-Z.}\ \bibnamefont {You}}, \ and\
  \bibinfo {author} {\bibfnamefont {C.}~\bibnamefont {Xu}},\ }\href {\doibase
  10.1103/PhysRevB.90.081110} {\bibfield  {journal} {\bibinfo  {journal} {Phys.
  Rev. B}\ }\textbf {\bibinfo {volume} {90}},\ \bibinfo {pages} {081110}
  (\bibinfo {year} {2014})}\BibitemShut {NoStop}%
\bibitem [{\citenamefont {Wang}\ and\ \citenamefont
  {Levin}(2015)}]{Wang2015Topological}%
  \BibitemOpen
  \bibfield  {author} {\bibinfo {author} {\bibfnamefont {C.}~\bibnamefont
  {Wang}}\ and\ \bibinfo {author} {\bibfnamefont {M.}~\bibnamefont {Levin}},\
  }\href {\doibase 10.1103/PhysRevB.91.165119} {\bibfield  {journal} {\bibinfo
  {journal} {Phys. Rev. B}\ }\textbf {\bibinfo {volume} {91}},\ \bibinfo
  {pages} {165119} (\bibinfo {year} {2015})}\BibitemShut {NoStop}%
\bibitem [{\citenamefont {Lin}\ and\ \citenamefont
  {Levin}(2015)}]{Lin2015Loop}%
  \BibitemOpen
  \bibfield  {author} {\bibinfo {author} {\bibfnamefont {C.-H.}\ \bibnamefont
  {Lin}}\ and\ \bibinfo {author} {\bibfnamefont {M.}~\bibnamefont {Levin}},\
  }\href {\doibase 10.1103/PhysRevB.92.035115} {\bibfield  {journal} {\bibinfo
  {journal} {Phys. Rev. B}\ }\textbf {\bibinfo {volume} {92}},\ \bibinfo
  {pages} {035115} (\bibinfo {year} {2015})}\BibitemShut {NoStop}%
\bibitem [{\citenamefont {Jian}\ and\ \citenamefont {Qi}(2014)}]{Jian2014}%
  \BibitemOpen
  \bibfield  {author} {\bibinfo {author} {\bibfnamefont {C.-M.}\ \bibnamefont
  {Jian}}\ and\ \bibinfo {author} {\bibfnamefont {X.-L.}\ \bibnamefont {Qi}},\
  }\href {\doibase 10.1103/PhysRevX.4.041043} {\bibfield  {journal} {\bibinfo
  {journal} {Phys. Rev. X}\ }\textbf {\bibinfo {volume} {4}},\ \bibinfo {pages}
  {041043} (\bibinfo {year} {2014})}\BibitemShut {NoStop}%
\bibitem [{\citenamefont {Yoshida}(2015)}]{Yoshida2015Topological}%
  \BibitemOpen
  \bibfield  {author} {\bibinfo {author} {\bibfnamefont {B.}~\bibnamefont
  {Yoshida}},\ }\href {\doibase 10.1103/PhysRevB.91.245131} {\bibfield
  {journal} {\bibinfo  {journal} {Phys. Rev. B}\ }\textbf {\bibinfo {volume}
  {91}},\ \bibinfo {pages} {245131} (\bibinfo {year} {2015})}\BibitemShut
  {NoStop}%
\bibitem [{\citenamefont {Chen}\ \emph {et~al.}(2016)\citenamefont {Chen},
  \citenamefont {Tiwari},\ and\ \citenamefont {Ryu}}]{Chen2016Bulk}%
  \BibitemOpen
  \bibfield  {author} {\bibinfo {author} {\bibfnamefont {X.}~\bibnamefont
  {Chen}}, \bibinfo {author} {\bibfnamefont {A.}~\bibnamefont {Tiwari}}, \ and\
  \bibinfo {author} {\bibfnamefont {S.}~\bibnamefont {Ryu}},\ }\href {\doibase
  10.1103/PhysRevB.94.045113} {\bibfield  {journal} {\bibinfo  {journal} {Phys.
  Rev. B}\ }\textbf {\bibinfo {volume} {94}},\ \bibinfo {pages} {045113}
  (\bibinfo {year} {2016})}\BibitemShut {NoStop}%
\bibitem [{\citenamefont {Wang}\ \emph {et~al.}(2016)\citenamefont {Wang},
  \citenamefont {Wen},\ and\ \citenamefont {Yau}}]{Wang2016Quantum}%
  \BibitemOpen
  \bibfield  {author} {\bibinfo {author} {\bibfnamefont {J.}~\bibnamefont
  {Wang}}, \bibinfo {author} {\bibfnamefont {X.-G.}\ \bibnamefont {Wen}}, \
  and\ \bibinfo {author} {\bibfnamefont {S.-T.}\ \bibnamefont {Yau}},\
  }\href@noop {} {\bibfield  {journal} {\bibinfo  {journal} {arXiv preprint
  arXiv:1602.05951}\ } (\bibinfo {year} {2016})}\BibitemShut {NoStop}%
\bibitem [{\citenamefont {Wang}(2016)}]{Wang2016Aspects}%
  \BibitemOpen
  \bibfield  {author} {\bibinfo {author} {\bibfnamefont {J.~C.}\ \bibnamefont
  {Wang}},\ }\href@noop {} {\bibfield  {journal} {\bibinfo  {journal} {arXiv
  preprint arXiv:1602.05569}\ } (\bibinfo {year} {2016})}\BibitemShut {NoStop}%
\bibitem [{\citenamefont {Tiwari}\ \emph
  {et~al.}(2017{\natexlab{a}})\citenamefont {Tiwari}, \citenamefont {Chen},\
  and\ \citenamefont {Ryu}}]{Tiwari2017Wilson}%
  \BibitemOpen
  \bibfield  {author} {\bibinfo {author} {\bibfnamefont {A.}~\bibnamefont
  {Tiwari}}, \bibinfo {author} {\bibfnamefont {X.}~\bibnamefont {Chen}}, \ and\
  \bibinfo {author} {\bibfnamefont {S.}~\bibnamefont {Ryu}},\ }\href {\doibase
  10.1103/PhysRevB.95.245124} {\bibfield  {journal} {\bibinfo  {journal} {Phys.
  Rev. B}\ }\textbf {\bibinfo {volume} {95}},\ \bibinfo {pages} {245124}
  (\bibinfo {year} {2017}{\natexlab{a}})}\BibitemShut {NoStop}%
\bibitem [{\citenamefont {Wang}\ and\ \citenamefont
  {Chen}(2017{\natexlab{a}})}]{Wang2017Twisted}%
  \BibitemOpen
  \bibfield  {author} {\bibinfo {author} {\bibfnamefont {Z.}~\bibnamefont
  {Wang}}\ and\ \bibinfo {author} {\bibfnamefont {X.}~\bibnamefont {Chen}},\
  }\href {\doibase 10.1103/PhysRevB.95.115142} {\bibfield  {journal} {\bibinfo
  {journal} {Phys. Rev. B}\ }\textbf {\bibinfo {volume} {95}},\ \bibinfo
  {pages} {115142} (\bibinfo {year} {2017}{\natexlab{a}})}\BibitemShut
  {NoStop}%
\bibitem [{\citenamefont {Williamson}\ and\ \citenamefont
  {Wang}(2017)}]{Williamson2017311}%
  \BibitemOpen
  \bibfield  {author} {\bibinfo {author} {\bibfnamefont {D.~J.}\ \bibnamefont
  {Williamson}}\ and\ \bibinfo {author} {\bibfnamefont {Z.}~\bibnamefont
  {Wang}},\ }\href {\doibase https://doi.org/10.1016/j.aop.2016.12.018}
  {\bibfield  {journal} {\bibinfo  {journal} {Annals of Physics}\ }\textbf
  {\bibinfo {volume} {377}},\ \bibinfo {pages} {311 } (\bibinfo {year}
  {2017})}\BibitemShut {NoStop}%
\bibitem [{\citenamefont {Chan}\ \emph {et~al.}(2017)\citenamefont {Chan},
  \citenamefont {Ye},\ and\ \citenamefont {Ryu}}]{Chan2017Borromean}%
  \BibitemOpen
  \bibfield  {author} {\bibinfo {author} {\bibfnamefont {A.~P.}\ \bibnamefont
  {Chan}}, \bibinfo {author} {\bibfnamefont {P.}~\bibnamefont {Ye}}, \ and\
  \bibinfo {author} {\bibfnamefont {S.}~\bibnamefont {Ryu}},\ }\href@noop {}
  {\bibfield  {journal} {\bibinfo  {journal} {arXiv preprint arXiv:1703.01926}\
  } (\bibinfo {year} {2017})}\BibitemShut {NoStop}%
\bibitem [{\citenamefont {{Cheng}}(2015)}]{Cheng1511}%
  \BibitemOpen
  \bibfield  {author} {\bibinfo {author} {\bibfnamefont {M.}~\bibnamefont
  {{Cheng}}},\ }\href@noop {} {\bibfield  {journal} {\bibinfo  {journal} {ArXiv
  e-prints}\ } (\bibinfo {year} {2015})},\ \Eprint
  {http://arxiv.org/abs/1511.02563} {arXiv:1511.02563 [cond-mat.str-el]}
  \BibitemShut {NoStop}%
\bibitem [{\citenamefont {Cheng}\ \emph {et~al.}(2017)\citenamefont {Cheng},
  \citenamefont {Tantivasadakarn},\ and\ \citenamefont {Wang}}]{Cheng2017Loop}%
  \BibitemOpen
  \bibfield  {author} {\bibinfo {author} {\bibfnamefont {M.}~\bibnamefont
  {Cheng}}, \bibinfo {author} {\bibfnamefont {N.}~\bibnamefont
  {Tantivasadakarn}}, \ and\ \bibinfo {author} {\bibfnamefont {C.}~\bibnamefont
  {Wang}},\ }\href@noop {} {\bibfield  {journal} {\bibinfo  {journal} {arXiv
  preprint arXiv:1705.08911}\ } (\bibinfo {year} {2017})}\BibitemShut {NoStop}%
\bibitem [{\citenamefont {Putrov}\ \emph {et~al.}(2017)\citenamefont {Putrov},
  \citenamefont {Wang},\ and\ \citenamefont {Yau}}]{Putrov2017}%
  \BibitemOpen
  \bibfield  {author} {\bibinfo {author} {\bibfnamefont {P.}~\bibnamefont
  {Putrov}}, \bibinfo {author} {\bibfnamefont {J.}~\bibnamefont {Wang}}, \ and\
  \bibinfo {author} {\bibfnamefont {S.-T.}\ \bibnamefont {Yau}},\ }\href
  {\doibase https://doi.org/10.1016/j.aop.2017.06.019} {\bibfield  {journal}
  {\bibinfo  {journal} {Annals of Physics}\ }\textbf {\bibinfo {volume}
  {384}},\ \bibinfo {pages} {254 } (\bibinfo {year} {2017})}\BibitemShut
  {NoStop}%
\bibitem [{\citenamefont {Wang}\ and\ \citenamefont
  {Chen}(2017{\natexlab{b}})}]{Chen2017}%
  \BibitemOpen
  \bibfield  {author} {\bibinfo {author} {\bibfnamefont {Z.}~\bibnamefont
  {Wang}}\ and\ \bibinfo {author} {\bibfnamefont {X.}~\bibnamefont {Chen}},\
  }\href {\doibase 10.1103/PhysRevB.95.115142} {\bibfield  {journal} {\bibinfo
  {journal} {Phys. Rev. B}\ }\textbf {\bibinfo {volume} {95}},\ \bibinfo
  {pages} {115142} (\bibinfo {year} {2017}{\natexlab{b}})}\BibitemShut
  {NoStop}%
\bibitem [{\citenamefont {Else}\ and\ \citenamefont {Nayak}(2017)}]{Else2017}%
  \BibitemOpen
  \bibfield  {author} {\bibinfo {author} {\bibfnamefont {D.~V.}\ \bibnamefont
  {Else}}\ and\ \bibinfo {author} {\bibfnamefont {C.}~\bibnamefont {Nayak}},\
  }\href {\doibase 10.1103/PhysRevB.96.045136} {\bibfield  {journal} {\bibinfo
  {journal} {Phys. Rev. B}\ }\textbf {\bibinfo {volume} {96}},\ \bibinfo
  {pages} {045136} (\bibinfo {year} {2017})}\BibitemShut {NoStop}%
\bibitem [{\citenamefont {Delcamp}(2017)}]{DelCamp1709}%
  \BibitemOpen
  \bibfield  {author} {\bibinfo {author} {\bibfnamefont {C.}~\bibnamefont
  {Delcamp}},\ }\href@noop {} {\bibfield  {journal} {\bibinfo  {journal}
  {arXiv: 1709.04924}\ } (\bibinfo {year} {2017})}\BibitemShut {NoStop}%
\bibitem [{\citenamefont {Levin}\ and\ \citenamefont
  {Gu}(2012)}]{Levin2012Braiding}%
  \BibitemOpen
  \bibfield  {author} {\bibinfo {author} {\bibfnamefont {M.}~\bibnamefont
  {Levin}}\ and\ \bibinfo {author} {\bibfnamefont {Z.-C.}\ \bibnamefont {Gu}},\
  }\href {\doibase 10.1103/PhysRevB.86.115109} {\bibfield  {journal} {\bibinfo
  {journal} {Phys. Rev. B}\ }\textbf {\bibinfo {volume} {86}},\ \bibinfo
  {pages} {115109} (\bibinfo {year} {2012})}\BibitemShut {NoStop}%
\bibitem [{\citenamefont {Ye}\ and\ \citenamefont {Wang}(2013)}]{Ye2013}%
  \BibitemOpen
  \bibfield  {author} {\bibinfo {author} {\bibfnamefont {P.}~\bibnamefont
  {Ye}}\ and\ \bibinfo {author} {\bibfnamefont {J.}~\bibnamefont {Wang}},\
  }\href {\doibase 10.1103/PhysRevB.88.235109} {\bibfield  {journal} {\bibinfo
  {journal} {Phys. Rev. B}\ }\textbf {\bibinfo {volume} {88}},\ \bibinfo
  {pages} {235109} (\bibinfo {year} {2013})}\BibitemShut {NoStop}%
\bibitem [{\citenamefont {Ye}\ and\ \citenamefont {Gu}(2016)}]{Ye2016}%
  \BibitemOpen
  \bibfield  {author} {\bibinfo {author} {\bibfnamefont {P.}~\bibnamefont
  {Ye}}\ and\ \bibinfo {author} {\bibfnamefont {Z.-C.}\ \bibnamefont {Gu}},\
  }\href {\doibase 10.1103/PhysRevB.93.205157} {\bibfield  {journal} {\bibinfo
  {journal} {Phys. Rev. B}\ }\textbf {\bibinfo {volume} {93}},\ \bibinfo
  {pages} {205157} (\bibinfo {year} {2016})}\BibitemShut {NoStop}%
\bibitem [{\citenamefont {He}\ \emph {et~al.}(2017)\citenamefont {He},
  \citenamefont {Zheng},\ and\ \citenamefont {von Keyserlingk}}]{He1608}%
  \BibitemOpen
  \bibfield  {author} {\bibinfo {author} {\bibfnamefont {H.}~\bibnamefont
  {He}}, \bibinfo {author} {\bibfnamefont {Y.}~\bibnamefont {Zheng}}, \ and\
  \bibinfo {author} {\bibfnamefont {C.}~\bibnamefont {von Keyserlingk}},\
  }\href {\doibase 10.1103/PhysRevB.95.035131} {\bibfield  {journal} {\bibinfo
  {journal} {Phys. Rev. B}\ }\textbf {\bibinfo {volume} {95}},\ \bibinfo
  {pages} {035131} (\bibinfo {year} {2017})}\BibitemShut {NoStop}%
\bibitem [{\citenamefont {Gu}\ and\ \citenamefont {Wen}(2009)}]{Gu2009Tensor}%
  \BibitemOpen
  \bibfield  {author} {\bibinfo {author} {\bibfnamefont {Z.-C.}\ \bibnamefont
  {Gu}}\ and\ \bibinfo {author} {\bibfnamefont {X.-G.}\ \bibnamefont {Wen}},\
  }\href {\doibase 10.1103/PhysRevB.80.155131} {\bibfield  {journal} {\bibinfo
  {journal} {Phys. Rev. B}\ }\textbf {\bibinfo {volume} {80}},\ \bibinfo
  {pages} {155131} (\bibinfo {year} {2009})}\BibitemShut {NoStop}%
\bibitem [{\citenamefont {Pollmann}\ \emph {et~al.}(2010)\citenamefont
  {Pollmann}, \citenamefont {Turner}, \citenamefont {Berg},\ and\ \citenamefont
  {Oshikawa}}]{Pollmann2010Entanglement}%
  \BibitemOpen
  \bibfield  {author} {\bibinfo {author} {\bibfnamefont {F.}~\bibnamefont
  {Pollmann}}, \bibinfo {author} {\bibfnamefont {A.~M.}\ \bibnamefont
  {Turner}}, \bibinfo {author} {\bibfnamefont {E.}~\bibnamefont {Berg}}, \ and\
  \bibinfo {author} {\bibfnamefont {M.}~\bibnamefont {Oshikawa}},\ }\href
  {\doibase 10.1103/PhysRevB.81.064439} {\bibfield  {journal} {\bibinfo
  {journal} {Phys. Rev. B}\ }\textbf {\bibinfo {volume} {81}},\ \bibinfo
  {pages} {064439} (\bibinfo {year} {2010})}\BibitemShut {NoStop}%
\bibitem [{\citenamefont {Chen}\ \emph
  {et~al.}(2011{\natexlab{a}})\citenamefont {Chen}, \citenamefont {Gu},\ and\
  \citenamefont {Wen}}]{Chen2011Classification}%
  \BibitemOpen
  \bibfield  {author} {\bibinfo {author} {\bibfnamefont {X.}~\bibnamefont
  {Chen}}, \bibinfo {author} {\bibfnamefont {Z.-C.}\ \bibnamefont {Gu}}, \ and\
  \bibinfo {author} {\bibfnamefont {X.-G.}\ \bibnamefont {Wen}},\ }\href
  {\doibase 10.1103/PhysRevB.83.035107} {\bibfield  {journal} {\bibinfo
  {journal} {Phys. Rev. B}\ }\textbf {\bibinfo {volume} {83}},\ \bibinfo
  {pages} {035107} (\bibinfo {year} {2011}{\natexlab{a}})}\BibitemShut
  {NoStop}%
\bibitem [{\citenamefont {Liu}\ \emph {et~al.}(2011{\natexlab{a}})\citenamefont
  {Liu}, \citenamefont {Liu},\ and\ \citenamefont {Wen}}]{Liu2011Gapped}%
  \BibitemOpen
  \bibfield  {author} {\bibinfo {author} {\bibfnamefont {Z.-X.}\ \bibnamefont
  {Liu}}, \bibinfo {author} {\bibfnamefont {M.}~\bibnamefont {Liu}}, \ and\
  \bibinfo {author} {\bibfnamefont {X.-G.}\ \bibnamefont {Wen}},\ }\href
  {\doibase 10.1103/PhysRevB.84.075135} {\bibfield  {journal} {\bibinfo
  {journal} {Phys. Rev. B}\ }\textbf {\bibinfo {volume} {84}},\ \bibinfo
  {pages} {075135} (\bibinfo {year} {2011}{\natexlab{a}})}\BibitemShut
  {NoStop}%
\bibitem [{\citenamefont {Chen}\ \emph
  {et~al.}(2011{\natexlab{b}})\citenamefont {Chen}, \citenamefont {Gu},\ and\
  \citenamefont {Wen}}]{Chen2011Complete}%
  \BibitemOpen
  \bibfield  {author} {\bibinfo {author} {\bibfnamefont {X.}~\bibnamefont
  {Chen}}, \bibinfo {author} {\bibfnamefont {Z.-C.}\ \bibnamefont {Gu}}, \ and\
  \bibinfo {author} {\bibfnamefont {X.-G.}\ \bibnamefont {Wen}},\ }\href
  {\doibase 10.1103/PhysRevB.84.235128} {\bibfield  {journal} {\bibinfo
  {journal} {Phys. Rev. B}\ }\textbf {\bibinfo {volume} {84}},\ \bibinfo
  {pages} {235128} (\bibinfo {year} {2011}{\natexlab{b}})}\BibitemShut
  {NoStop}%
\bibitem [{\citenamefont {Liu}\ \emph {et~al.}(2011{\natexlab{b}})\citenamefont
  {Liu}, \citenamefont {Chen},\ and\ \citenamefont {Wen}}]{Liu2011Symmetry}%
  \BibitemOpen
  \bibfield  {author} {\bibinfo {author} {\bibfnamefont {Z.-X.}\ \bibnamefont
  {Liu}}, \bibinfo {author} {\bibfnamefont {X.}~\bibnamefont {Chen}}, \ and\
  \bibinfo {author} {\bibfnamefont {X.-G.}\ \bibnamefont {Wen}},\ }\href
  {\doibase 10.1103/PhysRevB.84.195145} {\bibfield  {journal} {\bibinfo
  {journal} {Phys. Rev. B}\ }\textbf {\bibinfo {volume} {84}},\ \bibinfo
  {pages} {195145} (\bibinfo {year} {2011}{\natexlab{b}})}\BibitemShut
  {NoStop}%
\bibitem [{\citenamefont {Pollmann}\ \emph {et~al.}(2012)\citenamefont
  {Pollmann}, \citenamefont {Berg}, \citenamefont {Turner},\ and\ \citenamefont
  {Oshikawa}}]{Pollmann2012Symmetry}%
  \BibitemOpen
  \bibfield  {author} {\bibinfo {author} {\bibfnamefont {F.}~\bibnamefont
  {Pollmann}}, \bibinfo {author} {\bibfnamefont {E.}~\bibnamefont {Berg}},
  \bibinfo {author} {\bibfnamefont {A.~M.}\ \bibnamefont {Turner}}, \ and\
  \bibinfo {author} {\bibfnamefont {M.}~\bibnamefont {Oshikawa}},\ }\href
  {\doibase 10.1103/PhysRevB.85.075125} {\bibfield  {journal} {\bibinfo
  {journal} {Phys. Rev. B}\ }\textbf {\bibinfo {volume} {85}},\ \bibinfo
  {pages} {075125} (\bibinfo {year} {2012})}\BibitemShut {NoStop}%
\bibitem [{\citenamefont {Wen}(2012)}]{Wen2012Symmetry}%
  \BibitemOpen
  \bibfield  {author} {\bibinfo {author} {\bibfnamefont {X.-G.}\ \bibnamefont
  {Wen}},\ }\href {\doibase 10.1103/PhysRevB.85.085103} {\bibfield  {journal}
  {\bibinfo  {journal} {Phys. Rev. B}\ }\textbf {\bibinfo {volume} {85}},\
  \bibinfo {pages} {085103} (\bibinfo {year} {2012})}\BibitemShut {NoStop}%
\bibitem [{\citenamefont {Tang}\ and\ \citenamefont
  {Wen}(2012)}]{Tang2012Interacting}%
  \BibitemOpen
  \bibfield  {author} {\bibinfo {author} {\bibfnamefont {E.}~\bibnamefont
  {Tang}}\ and\ \bibinfo {author} {\bibfnamefont {X.-G.}\ \bibnamefont {Wen}},\
  }\href {\doibase 10.1103/PhysRevLett.109.096403} {\bibfield  {journal}
  {\bibinfo  {journal} {Phys. Rev. Lett.}\ }\textbf {\bibinfo {volume} {109}},\
  \bibinfo {pages} {096403} (\bibinfo {year} {2012})}\BibitemShut {NoStop}%
\bibitem [{\citenamefont {Liu}\ \emph {et~al.}(2012)\citenamefont {Liu},
  \citenamefont {Yang}, \citenamefont {Han}, \citenamefont {Yi},\ and\
  \citenamefont {Wen}}]{Liu2012Symmetry}%
  \BibitemOpen
  \bibfield  {author} {\bibinfo {author} {\bibfnamefont {Z.-X.}\ \bibnamefont
  {Liu}}, \bibinfo {author} {\bibfnamefont {Z.-B.}\ \bibnamefont {Yang}},
  \bibinfo {author} {\bibfnamefont {Y.-J.}\ \bibnamefont {Han}}, \bibinfo
  {author} {\bibfnamefont {W.}~\bibnamefont {Yi}}, \ and\ \bibinfo {author}
  {\bibfnamefont {X.-G.}\ \bibnamefont {Wen}},\ }\href {\doibase
  10.1103/PhysRevB.86.195122} {\bibfield  {journal} {\bibinfo  {journal} {Phys.
  Rev. B}\ }\textbf {\bibinfo {volume} {86}},\ \bibinfo {pages} {195122}
  (\bibinfo {year} {2012})}\BibitemShut {NoStop}%
\bibitem [{\citenamefont {Chen}\ and\ \citenamefont
  {Wen}(2012)}]{Chen2012Chiral}%
  \BibitemOpen
  \bibfield  {author} {\bibinfo {author} {\bibfnamefont {X.}~\bibnamefont
  {Chen}}\ and\ \bibinfo {author} {\bibfnamefont {X.-G.}\ \bibnamefont {Wen}},\
  }\href {\doibase 10.1103/PhysRevB.86.235135} {\bibfield  {journal} {\bibinfo
  {journal} {Phys. Rev. B}\ }\textbf {\bibinfo {volume} {86}},\ \bibinfo
  {pages} {235135} (\bibinfo {year} {2012})}\BibitemShut {NoStop}%
\bibitem [{\citenamefont {Chen}\ \emph {et~al.}(2012)\citenamefont {Chen},
  \citenamefont {Gu}, \citenamefont {Liu},\ and\ \citenamefont
  {Wen}}]{chen2012symmetry}%
  \BibitemOpen
  \bibfield  {author} {\bibinfo {author} {\bibfnamefont {X.}~\bibnamefont
  {Chen}}, \bibinfo {author} {\bibfnamefont {Z.-C.}\ \bibnamefont {Gu}},
  \bibinfo {author} {\bibfnamefont {Z.-X.}\ \bibnamefont {Liu}}, \ and\
  \bibinfo {author} {\bibfnamefont {X.-G.}\ \bibnamefont {Wen}},\ }\href@noop
  {} {\bibfield  {journal} {\bibinfo  {journal} {Science}\ }\textbf {\bibinfo
  {volume} {338}},\ \bibinfo {pages} {1604} (\bibinfo {year}
  {2012})}\BibitemShut {NoStop}%
\bibitem [{\citenamefont {Chen}\ \emph {et~al.}(2013)\citenamefont {Chen},
  \citenamefont {Gu}, \citenamefont {Liu},\ and\ \citenamefont
  {Wen}}]{ChenGuLiuWen}%
  \BibitemOpen
  \bibfield  {author} {\bibinfo {author} {\bibfnamefont {X.}~\bibnamefont
  {Chen}}, \bibinfo {author} {\bibfnamefont {Z.-C.}\ \bibnamefont {Gu}},
  \bibinfo {author} {\bibfnamefont {Z.-X.}\ \bibnamefont {Liu}}, \ and\
  \bibinfo {author} {\bibfnamefont {X.-G.}\ \bibnamefont {Wen}},\ }\href
  {\doibase 10.1103/PhysRevB.87.155114} {\bibfield  {journal} {\bibinfo
  {journal} {Phys. Rev. B}\ }\textbf {\bibinfo {volume} {87}},\ \bibinfo
  {pages} {155114} (\bibinfo {year} {2013})}\BibitemShut {NoStop}%
\bibitem [{\citenamefont {Liu}\ and\ \citenamefont
  {Wen}(2013)}]{Liu2013Symmetry}%
  \BibitemOpen
  \bibfield  {author} {\bibinfo {author} {\bibfnamefont {Z.-X.}\ \bibnamefont
  {Liu}}\ and\ \bibinfo {author} {\bibfnamefont {X.-G.}\ \bibnamefont {Wen}},\
  }\href {\doibase 10.1103/PhysRevLett.110.067205} {\bibfield  {journal}
  {\bibinfo  {journal} {Phys. Rev. Lett.}\ }\textbf {\bibinfo {volume} {110}},\
  \bibinfo {pages} {067205} (\bibinfo {year} {2013})}\BibitemShut {NoStop}%
\bibitem [{\citenamefont {Wen}(2013)}]{Wen2013Classification}%
  \BibitemOpen
  \bibfield  {author} {\bibinfo {author} {\bibfnamefont {X.-G.}\ \bibnamefont
  {Wen}},\ }\href {\doibase 10.1103/PhysRevD.88.045013} {\bibfield  {journal}
  {\bibinfo  {journal} {Phys. Rev. D}\ }\textbf {\bibinfo {volume} {88}},\
  \bibinfo {pages} {045013} (\bibinfo {year} {2013})}\BibitemShut {NoStop}%
\bibitem [{\citenamefont {Chen}\ \emph {et~al.}(2015)\citenamefont {Chen},
  \citenamefont {Burnell}, \citenamefont {Vishwanath},\ and\ \citenamefont
  {Fidkowski}}]{ChenBurnell2015}%
  \BibitemOpen
  \bibfield  {author} {\bibinfo {author} {\bibfnamefont {X.}~\bibnamefont
  {Chen}}, \bibinfo {author} {\bibfnamefont {F.~J.}\ \bibnamefont {Burnell}},
  \bibinfo {author} {\bibfnamefont {A.}~\bibnamefont {Vishwanath}}, \ and\
  \bibinfo {author} {\bibfnamefont {L.}~\bibnamefont {Fidkowski}},\ }\href
  {\doibase 10.1103/PhysRevX.5.041013} {\bibfield  {journal} {\bibinfo
  {journal} {Phys. Rev. X}\ }\textbf {\bibinfo {volume} {5}},\ \bibinfo {pages}
  {041013} (\bibinfo {year} {2015})}\BibitemShut {NoStop}%
\bibitem [{\citenamefont {Vishwanath}\ and\ \citenamefont
  {Senthil}(2013)}]{AshvinSenthil}%
  \BibitemOpen
  \bibfield  {author} {\bibinfo {author} {\bibfnamefont {A.}~\bibnamefont
  {Vishwanath}}\ and\ \bibinfo {author} {\bibfnamefont {T.}~\bibnamefont
  {Senthil}},\ }\href {\doibase 10.1103/PhysRevX.3.011016} {\bibfield
  {journal} {\bibinfo  {journal} {Phys. Rev. X}\ }\textbf {\bibinfo {volume}
  {3}},\ \bibinfo {pages} {011016} (\bibinfo {year} {2013})}\BibitemShut
  {NoStop}%
\bibitem [{\citenamefont {Wang}\ and\ \citenamefont
  {Senthil}(2014)}]{WangSenthil2013}%
  \BibitemOpen
  \bibfield  {author} {\bibinfo {author} {\bibfnamefont {C.}~\bibnamefont
  {Wang}}\ and\ \bibinfo {author} {\bibfnamefont {T.}~\bibnamefont {Senthil}},\
  }\href {\doibase 10.1103/PhysRevB.89.195124} {\bibfield  {journal} {\bibinfo
  {journal} {Phys. Rev. B}\ }\textbf {\bibinfo {volume} {89}},\ \bibinfo
  {pages} {195124} (\bibinfo {year} {2014})}\BibitemShut {NoStop}%
\bibitem [{\citenamefont {Wang}\ \emph {et~al.}(2014)\citenamefont {Wang},
  \citenamefont {Potter},\ and\ \citenamefont {Senthil}}]{Wang629}%
  \BibitemOpen
  \bibfield  {author} {\bibinfo {author} {\bibfnamefont {C.}~\bibnamefont
  {Wang}}, \bibinfo {author} {\bibfnamefont {A.~C.}\ \bibnamefont {Potter}}, \
  and\ \bibinfo {author} {\bibfnamefont {T.}~\bibnamefont {Senthil}},\ }\href
  {\doibase 10.1126/science.1243326} {\bibfield  {journal} {\bibinfo  {journal}
  {Science}\ }\textbf {\bibinfo {volume} {343}},\ \bibinfo {pages} {629}
  (\bibinfo {year} {2014})},\ \Eprint
  {http://arxiv.org/abs/http://science.sciencemag.org/content/343/6171/629.full.pdf}
  {http://science.sciencemag.org/content/343/6171/629.full.pdf} \BibitemShut
  {NoStop}%
\bibitem [{\citenamefont {Metlitski}\ \emph {et~al.}(2015)\citenamefont
  {Metlitski}, \citenamefont {Kane},\ and\ \citenamefont
  {Fisher}}]{Metlitski2015}%
  \BibitemOpen
  \bibfield  {author} {\bibinfo {author} {\bibfnamefont {M.~A.}\ \bibnamefont
  {Metlitski}}, \bibinfo {author} {\bibfnamefont {C.~L.}\ \bibnamefont {Kane}},
  \ and\ \bibinfo {author} {\bibfnamefont {M.~P.~A.}\ \bibnamefont {Fisher}},\
  }\href {\doibase 10.1103/PhysRevB.92.125111} {\bibfield  {journal} {\bibinfo
  {journal} {Phys. Rev. B}\ }\textbf {\bibinfo {volume} {92}},\ \bibinfo
  {pages} {125111} (\bibinfo {year} {2015})}\BibitemShut {NoStop}%
\bibitem [{\citenamefont {{Metlitski}}\ \emph {et~al.}(2014)\citenamefont
  {{Metlitski}}, \citenamefont {{Fidkowski}}, \citenamefont {{Chen}},\ and\
  \citenamefont {{Vishwanath}}}]{Metlitski2014arXiv}%
  \BibitemOpen
  \bibfield  {author} {\bibinfo {author} {\bibfnamefont {M.~A.}\ \bibnamefont
  {{Metlitski}}}, \bibinfo {author} {\bibfnamefont {L.}~\bibnamefont
  {{Fidkowski}}}, \bibinfo {author} {\bibfnamefont {X.}~\bibnamefont {{Chen}}},
  \ and\ \bibinfo {author} {\bibfnamefont {A.}~\bibnamefont {{Vishwanath}}},\
  }\href@noop {} {\bibfield  {journal} {\bibinfo  {journal} {ArXiv e-prints}\ }
  (\bibinfo {year} {2014})},\ \Eprint {http://arxiv.org/abs/1406.3032}
  {arXiv:1406.3032 [cond-mat.str-el]} \BibitemShut {NoStop}%
\bibitem [{\citenamefont {Gu}\ and\ \citenamefont
  {Wen}(2014)}]{Gu2014Symmetry}%
  \BibitemOpen
  \bibfield  {author} {\bibinfo {author} {\bibfnamefont {Z.-C.}\ \bibnamefont
  {Gu}}\ and\ \bibinfo {author} {\bibfnamefont {X.-G.}\ \bibnamefont {Wen}},\
  }\href {\doibase 10.1103/PhysRevB.90.115141} {\bibfield  {journal} {\bibinfo
  {journal} {Phys. Rev. B}\ }\textbf {\bibinfo {volume} {90}},\ \bibinfo
  {pages} {115141} (\bibinfo {year} {2014})}\BibitemShut {NoStop}%
\bibitem [{\citenamefont {Ye}\ and\ \citenamefont
  {Wen}(2014)}]{Ye2014Constructing}%
  \BibitemOpen
  \bibfield  {author} {\bibinfo {author} {\bibfnamefont {P.}~\bibnamefont
  {Ye}}\ and\ \bibinfo {author} {\bibfnamefont {X.-G.}\ \bibnamefont {Wen}},\
  }\href {\doibase 10.1103/PhysRevB.89.045127} {\bibfield  {journal} {\bibinfo
  {journal} {Phys. Rev. B}\ }\textbf {\bibinfo {volume} {89}},\ \bibinfo
  {pages} {045127} (\bibinfo {year} {2014})}\BibitemShut {NoStop}%
\bibitem [{\citenamefont {Liu}\ \emph {et~al.}(2014)\citenamefont {Liu},
  \citenamefont {Gu},\ and\ \citenamefont {Wen}}]{Liu2014Microscopic}%
  \BibitemOpen
  \bibfield  {author} {\bibinfo {author} {\bibfnamefont {Z.-X.}\ \bibnamefont
  {Liu}}, \bibinfo {author} {\bibfnamefont {Z.-C.}\ \bibnamefont {Gu}}, \ and\
  \bibinfo {author} {\bibfnamefont {X.-G.}\ \bibnamefont {Wen}},\ }\href
  {\doibase 10.1103/PhysRevLett.113.267206} {\bibfield  {journal} {\bibinfo
  {journal} {Phys. Rev. Lett.}\ }\textbf {\bibinfo {volume} {113}},\ \bibinfo
  {pages} {267206} (\bibinfo {year} {2014})}\BibitemShut {NoStop}%
\bibitem [{\citenamefont {Wen}(2015)}]{Wen2015Construction}%
  \BibitemOpen
  \bibfield  {author} {\bibinfo {author} {\bibfnamefont {X.-G.}\ \bibnamefont
  {Wen}},\ }\href {\doibase 10.1103/PhysRevB.91.205101} {\bibfield  {journal}
  {\bibinfo  {journal} {Phys. Rev. B}\ }\textbf {\bibinfo {volume} {91}},\
  \bibinfo {pages} {205101} (\bibinfo {year} {2015})}\BibitemShut {NoStop}%
\bibitem [{\citenamefont {Lan}\ \emph {et~al.}(2017)\citenamefont {Lan},
  \citenamefont {Kong},\ and\ \citenamefont {Wen}}]{Lan2017Classification}%
  \BibitemOpen
  \bibfield  {author} {\bibinfo {author} {\bibfnamefont {T.}~\bibnamefont
  {Lan}}, \bibinfo {author} {\bibfnamefont {L.}~\bibnamefont {Kong}}, \ and\
  \bibinfo {author} {\bibfnamefont {X.-G.}\ \bibnamefont {Wen}},\ }\href
  {\doibase 10.1103/PhysRevB.95.235140} {\bibfield  {journal} {\bibinfo
  {journal} {Phys. Rev. B}\ }\textbf {\bibinfo {volume} {95}},\ \bibinfo
  {pages} {235140} (\bibinfo {year} {2017})}\BibitemShut {NoStop}%
\bibitem [{\citenamefont {Tiwari}\ \emph
  {et~al.}(2017{\natexlab{b}})\citenamefont {Tiwari}, \citenamefont {Chen},
  \citenamefont {Shiozaki},\ and\ \citenamefont {Ryu}}]{Tiwari:2017wqf}%
  \BibitemOpen
  \bibfield  {author} {\bibinfo {author} {\bibfnamefont {A.}~\bibnamefont
  {Tiwari}}, \bibinfo {author} {\bibfnamefont {X.}~\bibnamefont {Chen}},
  \bibinfo {author} {\bibfnamefont {K.}~\bibnamefont {Shiozaki}}, \ and\
  \bibinfo {author} {\bibfnamefont {S.}~\bibnamefont {Ryu}},\ }\href@noop {} {\
   (\bibinfo {year} {2017}{\natexlab{b}})},\ \Eprint
  {http://arxiv.org/abs/1710.04730} {arXiv:1710.04730 [cond-mat.str-el]}
  \BibitemShut {NoStop}%
\bibitem [{\citenamefont {Dijkgraaf}\ and\ \citenamefont
  {Witten}(1990)}]{DW1990}%
  \BibitemOpen
  \bibfield  {author} {\bibinfo {author} {\bibfnamefont {R.}~\bibnamefont
  {Dijkgraaf}}\ and\ \bibinfo {author} {\bibfnamefont {E.}~\bibnamefont
  {Witten}},\ }\href@noop {} {\bibfield  {journal} {\bibinfo  {journal}
  {Communications in Mathematical Physics}\ }\textbf {\bibinfo {volume}
  {129}},\ \bibinfo {pages} {393} (\bibinfo {year} {1990})}\BibitemShut
  {NoStop}%
\bibitem [{\citenamefont {{Lan}}\ \emph {et~al.}(2017)\citenamefont {{Lan}},
  \citenamefont {{Kong}},\ and\ \citenamefont {{Wen}}}]{LanKongWen2017}%
  \BibitemOpen
  \bibfield  {author} {\bibinfo {author} {\bibfnamefont {T.}~\bibnamefont
  {{Lan}}}, \bibinfo {author} {\bibfnamefont {L.}~\bibnamefont {{Kong}}}, \
  and\ \bibinfo {author} {\bibfnamefont {X.-G.}\ \bibnamefont {{Wen}}},\
  }\href@noop {} {\bibfield  {journal} {\bibinfo  {journal} {ArXiv e-prints}\ }
  (\bibinfo {year} {2017})},\ \Eprint {http://arxiv.org/abs/1704.04221}
  {arXiv:1704.04221 [cond-mat.str-el]} \BibitemShut {NoStop}%
\bibitem [{\citenamefont {Grover}\ \emph {et~al.}(2011)\citenamefont {Grover},
  \citenamefont {Turner},\ and\ \citenamefont {Vishwanath}}]{Grover1108}%
  \BibitemOpen
  \bibfield  {author} {\bibinfo {author} {\bibfnamefont {T.}~\bibnamefont
  {Grover}}, \bibinfo {author} {\bibfnamefont {A.~M.}\ \bibnamefont {Turner}},
  \ and\ \bibinfo {author} {\bibfnamefont {A.}~\bibnamefont {Vishwanath}},\
  }\href {\doibase 10.1103/PhysRevB.84.195120} {\bibfield  {journal} {\bibinfo
  {journal} {Phys. Rev. B}\ }\textbf {\bibinfo {volume} {84}},\ \bibinfo
  {pages} {195120} (\bibinfo {year} {2011})}\BibitemShut {NoStop}%
\bibitem [{\citenamefont {{Zheng}}\ \emph {et~al.}(2017)\citenamefont
  {{Zheng}}, \citenamefont {{He}}, \citenamefont {{Bradlyn}}, \citenamefont
  {{Cano}}, \citenamefont {{Neupert}},\ and\ \citenamefont
  {{Bernevig}}}]{Zheng2017Structure}%
  \BibitemOpen
  \bibfield  {author} {\bibinfo {author} {\bibfnamefont {Y.}~\bibnamefont
  {{Zheng}}}, \bibinfo {author} {\bibfnamefont {H.}~\bibnamefont {{He}}},
  \bibinfo {author} {\bibfnamefont {B.}~\bibnamefont {{Bradlyn}}}, \bibinfo
  {author} {\bibfnamefont {J.}~\bibnamefont {{Cano}}}, \bibinfo {author}
  {\bibfnamefont {T.}~\bibnamefont {{Neupert}}}, \ and\ \bibinfo {author}
  {\bibfnamefont {B.~A.}\ \bibnamefont {{Bernevig}}},\ }\href@noop {}
  {\bibfield  {journal} {\bibinfo  {journal} {ArXiv e-prints}\ } (\bibinfo
  {year} {2017})},\ \Eprint {http://arxiv.org/abs/1710.01747} {arXiv:1710.01747
  [cond-mat.str-el]} \BibitemShut {NoStop}%
\bibitem [{\citenamefont {Hu}\ \emph {et~al.}(2013)\citenamefont {Hu},
  \citenamefont {Wan},\ and\ \citenamefont {Wu}}]{Hu2012}%
  \BibitemOpen
  \bibfield  {author} {\bibinfo {author} {\bibfnamefont {Y.}~\bibnamefont
  {Hu}}, \bibinfo {author} {\bibfnamefont {Y.}~\bibnamefont {Wan}}, \ and\
  \bibinfo {author} {\bibfnamefont {Y.-S.}\ \bibnamefont {Wu}},\ }\href
  {\doibase 10.1103/PhysRevB.87.125114} {\bibfield  {journal} {\bibinfo
  {journal} {Phys. Rev. B}\ }\textbf {\bibinfo {volume} {87}},\ \bibinfo
  {pages} {125114} (\bibinfo {year} {2013})}\BibitemShut {NoStop}%
\bibitem [{Note1()}]{Note1}%
  \BibitemOpen
  \bibinfo {note} {Mathematically, $\protect \mathrm {sRep}(G)$ is a category
  formed by the representations of $G$, with some of the irreducible
  representations assigned bosonic statistics, while the others assigned fermi
  statistics. $\protect \mathrm {Rep}(G)$ is a category formed by the
  representations of $G$, with all the irreducible representations assigned
  bosonic statistics. See, \protect \textit {e.g.}, Ref.\protect
  \rev@citealpnum {LanKongWen2016} for more details.}\BibitemShut {Stop}%
\bibitem [{\citenamefont {Preskill}\ and\ \citenamefont
  {Krauss}(1990)}]{Preskill199050}%
  \BibitemOpen
  \bibfield  {author} {\bibinfo {author} {\bibfnamefont {J.}~\bibnamefont
  {Preskill}}\ and\ \bibinfo {author} {\bibfnamefont {L.~M.}\ \bibnamefont
  {Krauss}},\ }\href {\doibase https://doi.org/10.1016/0550-3213(90)90262-C}
  {\bibfield  {journal} {\bibinfo  {journal} {Nuclear Physics B}\ }\textbf
  {\bibinfo {volume} {341}},\ \bibinfo {pages} {50 } (\bibinfo {year}
  {1990})}\BibitemShut {NoStop}%
\bibitem [{\citenamefont {Bucher}\ \emph {et~al.}(1992)\citenamefont {Bucher},
  \citenamefont {Lee},\ and\ \citenamefont {Preskill}}]{Preskill1992}%
  \BibitemOpen
  \bibfield  {author} {\bibinfo {author} {\bibfnamefont {M.}~\bibnamefont
  {Bucher}}, \bibinfo {author} {\bibfnamefont {K.-M.}\ \bibnamefont {Lee}}, \
  and\ \bibinfo {author} {\bibfnamefont {J.}~\bibnamefont {Preskill}},\ }\href
  {\doibase https://doi.org/10.1016/0550-3213(92)90174-A} {\bibfield  {journal}
  {\bibinfo  {journal} {Nuclear Physics B}\ }\textbf {\bibinfo {volume}
  {386}},\ \bibinfo {pages} {27 } (\bibinfo {year} {1992})}\BibitemShut
  {NoStop}%
\bibitem [{\citenamefont {Alford}\ \emph {et~al.}(1992)\citenamefont {Alford},
  \citenamefont {Lee}, \citenamefont {March-Russell},\ and\ \citenamefont
  {Preskill}}]{Preskill1992NPB}%
  \BibitemOpen
  \bibfield  {author} {\bibinfo {author} {\bibfnamefont {M.~G.}\ \bibnamefont
  {Alford}}, \bibinfo {author} {\bibfnamefont {K.-M.}\ \bibnamefont {Lee}},
  \bibinfo {author} {\bibfnamefont {J.}~\bibnamefont {March-Russell}}, \ and\
  \bibinfo {author} {\bibfnamefont {J.}~\bibnamefont {Preskill}},\ }\href
  {\doibase https://doi.org/10.1016/0550-3213(92)90468-Q} {\bibfield  {journal}
  {\bibinfo  {journal} {Nuclear Physics B}\ }\textbf {\bibinfo {volume}
  {384}},\ \bibinfo {pages} {251 } (\bibinfo {year} {1992})}\BibitemShut
  {NoStop}%
\bibitem [{Note2()}]{Note2}%
  \BibitemOpen
  \bibinfo {note} {We thank Chenjie Wang for pointing out this issue to
  us.}\BibitemShut {Stop}%
\bibitem [{Note3()}]{Note3}%
  \BibitemOpen
  \bibinfo {note} {As we will see later, even for nontrivial 4-cocycles, it is
  still possible that the projective representation reduces to a linear
  representation, as long as the induced 2-cocycles are trivial.}\BibitemShut
  {Stop}%
\bibitem [{\citenamefont {Atiyah}(1988)}]{Atiyah1988}%
  \BibitemOpen
  \bibfield  {author} {\bibinfo {author} {\bibfnamefont {M.}~\bibnamefont
  {Atiyah}},\ }\href {\doibase 10.1007/BF02698547} {\bibfield  {journal}
  {\bibinfo  {journal} {Publications Math{\'e}matiques de l'Institut des Hautes
  {\'E}tudes Scientifiques}\ }\textbf {\bibinfo {volume} {68}},\ \bibinfo
  {pages} {175} (\bibinfo {year} {1988})}\BibitemShut {NoStop}%
\bibitem [{\citenamefont {Freed}(1992)}]{Freed1992}%
  \BibitemOpen
  \bibfield  {author} {\bibinfo {author} {\bibfnamefont {D.}~\bibnamefont
  {Freed}},\ }\href@noop {} {\enquote {\bibinfo {title} {Lectures on
  topological quantum field theory},}\ } (\bibinfo {year} {1992})\BibitemShut
  {NoStop}%
\bibitem [{\citenamefont {Witten}(1989)}]{witten1989quantum}%
  \BibitemOpen
  \bibfield  {author} {\bibinfo {author} {\bibfnamefont {E.}~\bibnamefont
  {Witten}},\ }\href@noop {} {\bibfield  {journal} {\bibinfo  {journal}
  {Communications in Mathematical Physics}\ }\textbf {\bibinfo {volume}
  {121}},\ \bibinfo {pages} {351} (\bibinfo {year} {1989})}\BibitemShut
  {NoStop}%
\bibitem [{\citenamefont {Wakui}(1992)}]{wakui1992}%
  \BibitemOpen
  \bibfield  {author} {\bibinfo {author} {\bibfnamefont {M.}~\bibnamefont
  {Wakui}},\ }\href {https://projecteuclid.org:443/euclid.ojm/1200784084}
  {\bibfield  {journal} {\bibinfo  {journal} {Osaka J. Math.}\ }\textbf
  {\bibinfo {volume} {29}},\ \bibinfo {pages} {675} (\bibinfo {year}
  {1992})}\BibitemShut {NoStop}%
\bibitem [{Note4()}]{Note4}%
  \BibitemOpen
  \bibinfo {note} {Firstly since $W(\gamma )$ is evaluated on a map in a
  particular homotopy class $[\gamma ]: M\to BG$, it must by definition be
  insensitive to deformations that leave it in the same homotopy class. Then
  noticing that there are no non-trivial bundles on $S^3$, the map $\gamma $
  can be deformed such that its pullback to $M$ gives a $G$-coloring of a
  triangulation of $M=S^3\times S^{1}$ which only has non-trivial elements of
  $G$ along $S^1$ and identity everywhere else. Now we recall that in general a
  $G$-coloring is an assignment of four elements of $G$ to each 4-simplex.
  Clearly all 4-simplices of the above coloring have atleast one (actually
  three) trivial elements. Furthermore since one builds Dijkgraaf-Witten
  theories from normalized cocycles i.e those such that for $[\omega ]\in
  H_{\protect \text {group}}(G,U(1))$, we require $\omega (1,g_1,g_2,g_3)=1$.
  Therefore $W(\gamma )=1$ for a $G$-bundle on $S^3\times S^{1}$.}\BibitemShut
  {Stop}%
\bibitem [{\citenamefont {Propitius}(1995)}]{Propitius95}%
  \BibitemOpen
  \bibfield  {author} {\bibinfo {author} {\bibfnamefont {M.~d.~W.}\
  \bibnamefont {Propitius}},\ }\href@noop {} {\bibfield  {journal} {\bibinfo
  {journal} {arXiv preprint hep-th/9511195}\ } (\bibinfo {year}
  {1995})}\BibitemShut {NoStop}%
\bibitem [{\citenamefont {Kapustin}(2014)}]{kapustin2014bosonic}%
  \BibitemOpen
  \bibfield  {author} {\bibinfo {author} {\bibfnamefont {A.}~\bibnamefont
  {Kapustin}},\ }\href@noop {} {\bibfield  {journal} {\bibinfo  {journal}
  {arXiv preprint arXiv:1404.6659}\ } (\bibinfo {year} {2014})}\BibitemShut
  {NoStop}%
\bibitem [{Note5()}]{Note5}%
  \BibitemOpen
  \bibinfo {note} {We emphasize that here the entanglement cut is a $S^2$,
  while the entanglement cut considered in Sec.~\ref {Sec: 3dEE} is a $T^2$. It
  is our future work to investigate if the effect of topological excitations on
  entanglement entropy depends on the topology of entanglement cut or not, and
  if yes, how it depends on the entanglement cut.}\BibitemShut {Stop}%
\bibitem [{\citenamefont {Mesaros}\ and\ \citenamefont
  {Ran}(2013)}]{Mesaros1212}%
  \BibitemOpen
  \bibfield  {author} {\bibinfo {author} {\bibfnamefont {A.}~\bibnamefont
  {Mesaros}}\ and\ \bibinfo {author} {\bibfnamefont {Y.}~\bibnamefont {Ran}},\
  }\href {\doibase 10.1103/PhysRevB.87.155115} {\bibfield  {journal} {\bibinfo
  {journal} {Phys. Rev. B}\ }\textbf {\bibinfo {volume} {87}},\ \bibinfo
  {pages} {155115} (\bibinfo {year} {2013})}\BibitemShut {NoStop}%
\bibitem [{\citenamefont {Kitaev}(2003)}]{kitaev2003fault}%
  \BibitemOpen
  \bibfield  {author} {\bibinfo {author} {\bibfnamefont {A.~Y.}\ \bibnamefont
  {Kitaev}},\ }\href@noop {} {\bibfield  {journal} {\bibinfo  {journal} {Annals
  of Physics}\ }\textbf {\bibinfo {volume} {303}},\ \bibinfo {pages} {2}
  (\bibinfo {year} {2003})}\BibitemShut {NoStop}%
\bibitem [{\citenamefont {Alford}\ \emph {et~al.}(1990)\citenamefont {Alford},
  \citenamefont {Benson}, \citenamefont {Coleman}, \citenamefont
  {March-Russell},\ and\ \citenamefont {Wilczek}}]{Alford1990}%
  \BibitemOpen
  \bibfield  {author} {\bibinfo {author} {\bibfnamefont {M.~G.}\ \bibnamefont
  {Alford}}, \bibinfo {author} {\bibfnamefont {K.}~\bibnamefont {Benson}},
  \bibinfo {author} {\bibfnamefont {S.}~\bibnamefont {Coleman}}, \bibinfo
  {author} {\bibfnamefont {J.}~\bibnamefont {March-Russell}}, \ and\ \bibinfo
  {author} {\bibfnamefont {F.}~\bibnamefont {Wilczek}},\ }\href {\doibase
  10.1103/PhysRevLett.64.1632} {\bibfield  {journal} {\bibinfo  {journal}
  {Phys. Rev. Lett.}\ }\textbf {\bibinfo {volume} {64}},\ \bibinfo {pages}
  {1632} (\bibinfo {year} {1990})}\BibitemShut {NoStop}%
\bibitem [{Note6()}]{Note6}%
  \BibitemOpen
  \bibinfo {note} {It is noted that not all Hopf-link loop excitations have
  higher quantum dimensions. For example, for nonvanishing flux $a=b$,
  Eq.~\protect \textup {\hbox {\mathsurround \z@ \protect \normalfont
  (\ignorespaces \ref {TrivialCondition}\unskip \@@italiccorr )}} still holds.
  Then the 2-cocycle $\beta _{a,b}$ is trivial, and one still has one
  dimensional quantum dimension.}\BibitemShut {Stop}%
\bibitem [{\citenamefont {Wang}\ \emph {et~al.}(2018)\citenamefont {Wang},
  \citenamefont {Ohmori}, \citenamefont {Putrov}, \citenamefont {Zheng},
  \citenamefont {Lin}, \citenamefont {Gao},\ and\ \citenamefont
  {Yau}}]{WangJ1801}%
  \BibitemOpen
  \bibfield  {author} {\bibinfo {author} {\bibfnamefont {J.}~\bibnamefont
  {Wang}}, \bibinfo {author} {\bibfnamefont {K.}~\bibnamefont {Ohmori}},
  \bibinfo {author} {\bibfnamefont {P.}~\bibnamefont {Putrov}}, \bibinfo
  {author} {\bibfnamefont {Y.}~\bibnamefont {Zheng}}, \bibinfo {author}
  {\bibfnamefont {H.}~\bibnamefont {Lin}}, \bibinfo {author} {\bibfnamefont
  {P.}~\bibnamefont {Gao}}, \ and\ \bibinfo {author} {\bibfnamefont {S.-T.}\
  \bibnamefont {Yau}},\ }\href@noop {} {\  (\bibinfo {year} {2018})},\ \Eprint
  {http://arxiv.org/abs/1801.05416} {arXiv:1801.05416 [cond-mat.str-el]}
  \BibitemShut {NoStop}%
\bibitem [{\citenamefont {Lan}\ \emph {et~al.}(2016)\citenamefont {Lan},
  \citenamefont {Kong},\ and\ \citenamefont {Wen}}]{LanKongWen2016}%
  \BibitemOpen
  \bibfield  {author} {\bibinfo {author} {\bibfnamefont {T.}~\bibnamefont
  {Lan}}, \bibinfo {author} {\bibfnamefont {L.}~\bibnamefont {Kong}}, \ and\
  \bibinfo {author} {\bibfnamefont {X.-G.}\ \bibnamefont {Wen}},\ }\href
  {\doibase 10.1103/PhysRevB.94.155113} {\bibfield  {journal} {\bibinfo
  {journal} {Phys. Rev. B}\ }\textbf {\bibinfo {volume} {94}},\ \bibinfo
  {pages} {155113} (\bibinfo {year} {2016})}\BibitemShut {NoStop}%
\end{thebibliography}%
\end{document}